\documentclass[twocolumn,aps,pre,reprint,superscriptaddress,nofootinbib]{revtex4-1}
\usepackage{graphicx}
\usepackage{bm}
\usepackage{mathrsfs}
\usepackage{amsmath}
\usepackage{verbatim}
\usepackage[all, knot]{xy}
\xyoption{arc}

\usepackage{cancel}
\usepackage{dcolumn}
\usepackage{hyperref}
\usepackage{verbatim}
\usepackage{subfigure}  
\begin{document}


\title{Poincare recurrence and spectral cascades in 3D quantum turbulence}

\author{George Vahala${}^{1}$, Jeffrey Yepez${}^{2}$, Linda Vahala${}^{3}$, Min Soe${}^{4}$, Bo Zhang${}^1$, and Sean Ziegeler${}^5$
}


\address{
${}^{1}$Department of Physics, William \& Mary, Williamsburg, VA 23185\\
${}^{2}$Air Force Research Laboratory, Hanscom Air Force Base, MA  01731\\
${}^{3}$Department of Electrical \& Computer Engineering, Old Dominion University, Norfolk, VA 23529\\
${}^{4}$Department of Mathematics  and Physical Sciences, Rogers State University, Claremore, OK 74017\\
${}^{5}$High Performance Technologies, Inc., Reston, VA 20190
}

\begin{abstract}
The time evolution of the ground state wave function of a zero-temperature Bose-Einstein condensate (BEC) is well described by the Hamiltonian Gross-Pitaevskii (GP) equation.   Using a set of appropriately interleaved unitary collision-stream operators, a quantum lattice gas algorithm is devised which on taking moments recovers the Gross-Pitaevskii (GP) equation in the diffusion ordering (time scales as length$^2$).  Unexpectedly, there is a class of initial conditions in which their Poincare recurrence is extremely short.  As expected, the Poincare recurrence time scales with diffusion ordering as the grid is increased.   The spectral results of Yepez et. al. \cite{yepez2009prl} for quantum turbulence are revised and it is found that it is the \emph {compressible} kinetic energy spectrum that exhibits  3 strong cascade regions: a small-$k$ classical Kolmogorov $k^{-5/3}$, a steep semi-classical cascade region, and a large- $k$ quantum vortex spectrum $k^{-3}$.  For most evolution times the \emph{incompressible} kinetic energy spectrum exhibits a somewhat robust quantum vortex spectrum of $k^{-3}$ for an extended range in $k$ with a $k^{-3.38}$ spectrum for intermediate $k$.  For linear vortices of winding number 1 there is an intermittent loss of the quantum vortex cascade with its signature seen in the time evolution of the kinetic energy $E_{kin}(t)$, the loss of the quantum vortex spectrum $k^{-3}$ spectrum in the incompressible kinetic energy spectrum as well as the minimilization of the vortex core isosurfaces that would totally inhibit the Kelvin wave vortex cascade.  In the time intervals around these intermittencies the incompressible kinetic energy also exhibits a multi-cascade spectrum.
\end{abstract}


\maketitle

\section{Introduction}
 
The ground state of the many body wave function of a zero-temperature BEC is well described by the single particle wave function  $\varphi$ since all the bosons are in the same state.  Upon appropriate normalization, the evolution of this single particle wave function is governed by the Gross-Pitaevskii (GP) equation \cite{JMathPhys.1963.4.195,JETP.1961.2.451}
\begin{subequations}
\label{scalar_GP_forms}
\begin{equation}
\label{scalar_GP_standard_form}
i\hbar \partial_t \varphi =  -\frac{\hbar^2}{2m} \nabla^2 \varphi + ( g |\varphi|^2 - \mu) \varphi,
\end{equation}
for a spinless condensate where $g$ is the nonlinear coupling representing the $s-$wave scattering strength of the weak bosonic interactions in the mean-field approximation and $\mu$ is the chemical potential.  For numerical purposes it is convenient to rewrite (\ref{scalar_GP_standard_form}) as 
\begin{equation}
\label{scalar_GP}
i\partial_t \varphi =  - \nabla^2 \varphi + a( g |\varphi|^2 - 1) \varphi,
\end{equation}
\end{subequations}
where we consider a condensate comprised of particles of mass $m = 1/2$ in natural units where $\hbar=1$ and where $a$ is introduced as a radial scaling parameter (for tuning the numerical resolution of the vortex cores).  The ``single particle" wave function $\varphi$ is also the order parameter for the scalar BEC with $|\varphi| \rightarrow 0$ at each quantum vortex core (a topological singularity).

The GP system is Hamiltonian.  Now it is well known that for every Hamiltonian system there is a Poincar\'e recurrence of the initial conditions, in that the dynamics will eventually return arbitrarily close to their initial state---provided the dynamics are followed for a sufficiently long time \cite{ott2002}.  For nearly all continuous Hamiltonian systems, this Poincar\'e time is, however, effectively infinite.  Here we will show that there exists a special class of initial conditions for which this Poincare recurrence time is remarkably short and accessible with present day supercomputer resources.  To define this class we follow Nore et. al., \cite{nore1997physfluid} and split the total conserved energy into several components
 \begin{equation}
 \label{Energy_conservation}
 E_\text{TOT} = E_\text{kin}(t) + E_\text{qu}(t) + E_ \text{int}(t) = const.
 \end{equation} 
where the kinetic energy $E_\text{kin}(t)$, quantum energy $E_\text{qu}(t)$ and internal energy $E_\text{int}(t)$ are defined by
 \begin{equation}
 \label{Kinetic_Energy}
\begin{split}
 E_\text{kin}(t) = \int d^3\bm{x} [\sqrt{\rho}  \bm{v}]^2,
 &
 \qquad
 E_\text{qu}(t) = \frac{2}{a^2} \int d^3\bm{x} [\nabla \sqrt{\rho}]^2,
\\
  E_\text{int}(t) & = \frac{g}{a^2} \int d^3\bm{x} [\rho]^2,
\end{split}
 \end{equation} 
on using the Madelung transformation
\begin{equation}
\label{Madelung}
 \varphi =  \sqrt\rho \, e^{i \theta /2},   \qquad\text{with} \qquad  \bm{v} = \nabla \theta
\end{equation}
which relates the condensate wave function $\varphi$ to the BEC fluid density $\rho$ and velocity $\bm{v}$.  We shall find that for initial conditions with
\begin{equation}
\label{Poincare_loss}
 E_ \text{int}(0) \ll E_ \text{kin}(0), E_ \text{qu}(0)
 \end{equation} 
  the corresponding Poincar\'e recurrence time is extremely short.
  
  We have performed quantum turbulence simulations \cite{nore1997physfluid,feynmann1955,barenghi2008,tsubota2007} on the GP equation on grids of $5760^3$ \cite{yepez2009prl} and reported on the existence of 3 distinct energy cascade regions.
 Due to an oversight in these earlier computations, we attributed this triple cascade to the incompressible kinetic energy spectrum when, in fact, it should be basically attributed to the \emph{total} (and hence the \emph{compressible} ) kinetic energy spectrum.  This is verified by simulations on lattice grids of $3072^3$ with details presented in the Appendix.
In these new simulations one sees that the \emph{compressible} kinetic energy spectrum exhibits $3$ distinct spectral regions: a small $k$ classical Kolmogorov-like $k^{-5/3}$ spectrum, a steep semi-classical $k^{-\alpha}$ spectrum with a non-universal exponent $\alpha > 6$, followed by a robust quantum energy spectral region with $k^{-3}$ for large k.  The \emph{incompressible} kinetic is more complex.  There is a very robust $k^{-3}$ spectrum for a long wave number range extending to the maximally resolved $k$ in the simulation.  However, around the wave number $k$ where there is a switch from the semi-classical to quantum vortex energy spectrum in the \emph{compressible} spectrum, the \emph{incompressible} kinetic energy spectrum does exhibit a well-defined $k$ range in which the spectrum is $k^{-17/5}$.

The identification of the $k^{-3}$ and $k^{-17/5}$ spectra is strongly debated.  Since we have many quantum vortices present we will call the large-$k$ spectrum the quantum vortex spectrum.  As Nore et. al. \cite{nore1997physfluid} have pointed out, the energy spectrum of a single stationary linear quantum vortex is readily determined analytically and it exhibits only one long single spectrum of $k^{-3}$ all the way to the maximum $k$.  It is readily shown that this holds for both a 2D and a 3D single vortex.  
For either a Taylor-Green or Berloff vortex, all the kinetic energy is incompressible -- there is no compressible energy spectrum.  Of course, a $k^{-3}$ incompressible kinetic energy spectrum that is found in the dynamical evolution of the GP wave function, does not necessarily imply a simple quantum vortex.  The $k^{-17/5}$ incompressible kinetic energy spectrum is also quite interesting.  There is also a strong debate on the spectrum of quantum Kelvin waves on a quantum vortex.  Kozik \& Svistunov \cite{PhysRevB.77.060502} argue for a $k^{-17/5}$ spectrum in the \emph{wave-action} spectrum.  It has been shown \cite{proment} that in some regimes, the wave-action spectrum will have the same exponent as the kinetic energy spectra, and so it is not impossible that the $k^{-17/5}$ spectrum we observe could be associated with the quantum Kelvin wave cascade.  As regards the full kinetic energy spectrum, one can identify a classical Kolmogorov $k^{-5/3}$ spectrum for small k.  It is interesting to note that in subgrid scale turbulence closure modeling of sonic jet flows in compressible classical turbulence, it is deemed critical  \cite{genin2010} that the closure model reproduce the Kolmogorov scaling for the total kinetic energy.

In most of the simulations considered here we run on grids of $1200^3$ since the similar physics (Poincare recurrence and the associated phenomena) on a $3072^3$ would require 6.6 times more iterations.  What we shall find is a robust $k^{-3}$ \emph{incompressible} kinetic energy which is intermittently destroyed together with a minimization in the vortex core isosurfaces.  Away from these intermittencies one sees a dual kinetic energy spectrum. The total kinetic energy spectrum exhibits three different spectral regions - although somewhat weaker than the three spectral regions in the $3072^3$ simulations. 

In Section 2 we describe our unitary quantum lattice gas (QLG) algorithm which is extremely well parallelized on supercomputers (no performance saturation seen with scaling runs tested to 
$216 000$ cores )
and which will run on quantum computers when available.  Moreover QLG also has a low memory footprint.  In Section 3 we discuss the Poincare recurrence for a set of initial conditions satisfying Eq. (5).  The signature of this recurrence  is readily apparent on plotting the time evolution of the kinetic energy $E_\text{kin}(t)$ and quantum energy $E_\text{qu}(t)$.  Initially the rectilinear vortices are very well localized.  As a result, for nearly all regions in coordinate space the BEC density $\rho$ is constant.  From Eq. (3) this translates into very low initial quantum kinetic energy $E_\text{qu}(0) \ll E_\text{kin}(0)$ with $E_\text{comp}(0) \ll E_\text{incomp}(0)$ .  The Poincare recurrence is further established by looking at the vortex core isosurfaces.  These simulations are run on grids of $1200^{3}$.  In the continuum limit that yields Eq. (1), our QLG algorithm follows diffusion ordering.  This has been verified by increasing the grid from $L_{1}^{3}$ to $L_{2}^{3}$ and seeing the Poincare recurrence time increase from $T_{Poin}$ to $(L_{2}/L_{1})^{2} T_{Poin}$ (grids used were $512^3$,  $960^{3}$,  $1024^3$ and $1200^3$).  In Section 4 we shall discuss the somewhat unexpected intermittent loss of the quantum vortex spectrum in these Poincare recurrence simulations.  We find on examining the incompressible kinetic energy spectrum that the pronounced $k^{-3}$ spectrum is destroyed for approximately 2000 time steps.  A precursor for the loss of this vortex cascade is identified in the time evolution of the kinetic energy  $E_\text{kin}(t)$.  In Section 4 we also see how the topological changes in the quantum vortex cores (during this intermittent loss of energy spectrum) are also registered in the minimization of the vortex core isosurfaces.  Clearly, this minimization will preclude a quantum Kelvin wave cascade from occurring along the vortex core.  It is interesting to note 
\cite{bozhang2010} 
that in 2D GP quantum simulations, where the quantum vortex cores are just point singularities, one also sees the $k^{-3}$ large-k incompressible kinetic energy spectrum provided there were point vortex cores in the dynamics.  When the point vortices are annihilated during the GP dynamics, this $k^{-3}$ spectrum is lost.


  We here briefly put our QLG method into the context of the mesoscopic literature.  In computer simulations of nonlinear physics, it is desirable to develop algorithms that scale strongly with increasing number of processors.  Unfortunately, nearly all direct solution methods do not scale well because of the non-local nature of the equations.  Lattice Boltzmann and lattice gas algorithms \cite{succi2001} move to a mesoscopic level by either solving a simplified discretized collisional Boltzmann equation or by a simplified particle collide-stream set of rules on a lattice.  Constraints are imposed so that in the macroscopic limit one recovers the original system of equations.  What is gained is a very local representation that is ideally parallelized.  One of the earliest quantum lattice Boltzmann schemes was by Succi and Benzi \cite{succibenzi1993}.  Succi  \cite{succi1998} then extended his complex distribution function approach to examine the GP equation and extended the algorithm to 2D and 3D \cite{palpacelli2007, palpacelli2008}.  A quantum lattice gas approach based on qubits, similar to what is proposed here, was initiated by Yepez \cite{yepez1998intlmod,yepez2001pre} and extended by Yepez and Boghosian \cite{yepezboghosian2002}.  Quantum lattice algorithms were also examined by Meyer \cite{meyer2002}.  As our qubit lattice representation is a mesoscopic representation of the GP equation it is critical that it be benchmarked against exact solutions.  In particular, our codes have been benchmarked against the 1D soliton-soliton collisions of NLS in which the solitons retain their exact shape and speed together with a phase-induced shift due to the collision \cite{vahala2003}.  We have also benchmarked our codes for 1D vector soliton-soliton collisions in which one can have inelastic collisions for very specifically chosen initial amplitudes \cite{vahala2004,vahala2006}.  Finally we comment on the Poincare recurrence time of 1D NLS.  At first glance it would seem that the Poincare recurrence time for such a continuum Hamiltonian system would be effectively infinite.  However there were some hints in Tracy et. al. \cite{tracy1984} that the solution manifold for 1D NLS has an underlying finite-dimensional structure.  Thus, under certain initial conditions, one could find short Poincare recurrence times for 1D NLS.  What is more unexpected is that we have found a class of initial conditions which yield short Poincare recurrence for 3D quantum turbulence.  Outside this class, the Poincare recurrence time is essentially infinite.
  
 
\section{Quantum Lattice Gas Algorithm for the GP Equation}

To recover the scalar GP equation in the continuum limit, we consider two qubits at each lattice site.  Of the four possible states $|00\rangle$, $|01\rangle$, $|10\rangle$ and $|11\rangle$, we need consider just the complex amplitudes $\alpha$ and $\beta$ for the states $|01\rangle$, $|10\rangle$ respectively.  At each position $\bm{x}$ of the cubic lattice we introduce the two-spinor 
 \begin{equation}
\label{psi_2_component_form}
\psi(\bm{x},t) = \begin{pmatrix}
\alpha(\bm{x},t)   \\
 \beta(\bm{x},t)  
\end{pmatrix}
\end{equation}
and construct the evolution operator $U[\Omega]$ - consisting of an appropriate sequence of non-commuting unitary collision and streaming operators - so that in the continuum limit the two spinor equation
 \begin{equation}
\label{basic_unitary_evolution_equation}
\psi(\bm{x}, t+\Delta t) =  U[\Omega] \,\psi(\bm{x}, t).
\end{equation}
will reduce to the GP equation for the 1-particle boson wave function $\varphi$ under the projection
\begin{equation}
 (1 , 1) \cdot \psi  = \varphi 
\end{equation}

The unitary collision operator $C$ that entangles locally the amplitudes $\alpha$ and $\beta$ is chosen to be the square-root-of swap:
 \begin{equation}
\label{collide_operator}
 C \equiv e^{i   \frac{\pi}{4} \sigma_x (1-\sigma_x)} = \begin{pmatrix}
 \frac{1-i}{2}    & \frac{1+i}{2}   \\
   \frac{1+i}{2}   &  \frac{1-i}{2}
\end{pmatrix},
  \end{equation}
where the $\bm{\sigma}$ are the Pauli spin matrices 
\begin{equation}
\sigma_x =\begin{pmatrix}
   0   & 1   \\
   1   &  0
\end{pmatrix}
\qquad 
\sigma_y =\begin{pmatrix}
   0   & -i   \\
   i   &  0
\end{pmatrix}
\qquad 
\sigma_z =\begin{pmatrix}
   1   & 0   \\
   0   &  -1
\end{pmatrix}.
\end{equation}
$C^2$ is just the swap gate since
 \begin{equation}
C^{2} \begin{pmatrix}
\alpha   \\
 \beta  
\end{pmatrix} 
=
 \begin{pmatrix}
\beta  \\
 \alpha 
\end{pmatrix} .
\end{equation}

The streaming operators shift just one of these amplitudes at $\bm{x}$ to a neighboring lattice point  at $\bm{x} + \Delta \bm{x} $:
\begin{subequations}
 \begin{eqnarray}
S_{\Delta\bm{x}, 0} \begin{pmatrix}
\alpha(\bm{x},t)   \\
 \beta(\bm{x},t)  
\end{pmatrix}
&\equiv&
 \begin{pmatrix}
\alpha(\bm{x}+\Delta\bm{x},t)   \\
 \beta(\bm{x},t)  
\end{pmatrix}
\\
S_{\Delta\bm{x}, 1} \begin{pmatrix}
\alpha(\bm{x},t)   \\
 \beta(\bm{x},t)  
\end{pmatrix}
&\equiv&
 \begin{pmatrix}
\alpha(\bm{x},t)   \\
 \beta(\bm{x}+\Delta\bm{x},t)  
\end{pmatrix}.
\end{eqnarray}
\end{subequations}
The subscript $\gamma = 0$ on the streaming operator $S_{\Delta\bm{x}, \gamma}$ refers to shifting the amplitude $\alpha$ while the subscript $\gamma =1$ refers to shifting the amplitude $\beta$. 
In terms of the Paul spin matrices, the streaming operators can be written in the form
%
\begin{equation}
\label{stream_operators}
S_{\Delta\bm{x}, 0} = n + e^{\Delta \bm{x} \partial_{\bm{x}}}\,\bar n,
\qquad
 S_{\Delta\bm{x}, 1} =\bar n + e^{\Delta \bm{x} \partial_{\bm{x}}} \,n,
 \end{equation}
%
where $n=(1-\sigma_z)/2$, $\bar n=(1+\sigma_z)/2$.   It should be noted that the collision and streaming operators do not commute:  $[C,S]  \neq 0$.
	
We now consider the following interleaved sequence of unitary collision and streaming operators 
\begin{equation}
\label{interleaved_operator}
J_{x\gamma} =  S_{-\Delta \bm{x},\gamma}  C S_{\Delta \bm{x},\gamma}  C
\end{equation}
Since $|\Delta \bm{x}| \ll 1$ and  $C^4 = I$, $J_{x\gamma}^2 = I + O(\Delta \bm{x})$, where $I$ is the identity operator.  We first consider the effect of the evolution operator $U_ \gamma[\Omega(\bm{x})]$ 
 \begin{equation}
\label{basic_typeII_quantum_algorithm}
  U_ \gamma[\Omega(\bm{x})]= J_{x\gamma}^2 J_{y \gamma}^2 J_{z \gamma}^2 e^{-i  \varepsilon^2\Omega(\bm{x})} ,
\end{equation}
acting on the $\gamma$ component of the 2-spinor $\psi$.  Here $\varepsilon \ll 1$ is a perturbative parameter and $\Omega$ is a function to be specified later.

Using perturbation theory, it  can be shown that the time advancement of $\psi$
 \begin{equation}
\label{basic_unitary_evolution_equation}
\psi(\bm{x}, t+\Delta t) =  U_ \gamma[\Omega] \,\psi(\bm{x}, t).
\end{equation}
yields
\begin{equation}
\label{quantum_lattice_gas_equation_spinor_form}
\begin{split}
\psi(\bm{x}, t+\Delta t) 
&  = 
\psi(\bm{x}, t)
-i \varepsilon^2 \left[
-\frac{1}{2}\sigma_x\nabla^2
+\Omega 
\right]
\psi(\bm{x}, t)
+
\\
&
\frac{(-1)^\gamma\varepsilon^3}{4}(\sigma_y+\sigma_z)\nabla^3
\psi(\bm{x}, t)
+
{\cal O}(\varepsilon^4),
\end{split}
\end{equation}
 with $\gamma=0$ or $1$ and $\Delta \bm{x} = O(\varepsilon)$.  Since the order $\varepsilon^3$ term in (\ref{quantum_lattice_gas_equation_spinor_form}) changes sign with $\gamma$, one can eliminate this term by introducing the symmetrized evolution operator 
%
%
\begin{equation}
\label{symmetrized_evolution}
U[\Omega] = U_{1}\left[\frac{\Omega}{2}\right]U_{0}\left[\frac{\Omega}{2}\right].
\end{equation}
%
%
rather than just $U_{\gamma}$.  

Under diffusion ordering, $\Delta t = O(\varepsilon^2)$ and $\Delta \bm{x} = O(\varepsilon)$, the evolution equation
\begin{equation}
\psi(\bm{x},t+\Delta t) =  U[\Omega(\bm{x})] \,\psi(\bm{x}, t)
\end{equation}
 leads to a representation of the spinor equation
\begin{equation}
i \partial_t \psi(\bm{x}, t) = 
\left[
-\frac{1}{2}\sigma_x\nabla^2
+\Omega 
\right]
\psi(\bm{x}, t)
+
{\cal O}(\varepsilon^2),
\qquad
\end{equation}
where the function $\Omega$ is still arbitrary.  To recover the scalar GP equation, one simply rescales the spatial grid $\nabla \rightarrow a^{-1} \nabla$, contracts the 2-component field $\psi$ to the (scalar) BEC wave function $\varphi$
 \begin{equation}
\varphi = (1,1) \cdot \psi  = \alpha + \beta
\end{equation}
and chooses $\Omega = g |\varphi|^2-1$ :
\begin{equation}
\label{Gross_Pitaevskii_equation}
i \partial_t \varphi = - \nabla^2 \varphi +a (g| \varphi |^2 -1)\varphi\,
+
{\cal O}(\varepsilon^2).
\end{equation} 

One of the beauties of the QLG algorithm, Eqs. (20), (19), (16) and (15), is its ideal parallelization on current day supercomputers (as well its direct application to  quantum computers - once available - due to its unitary representation).  Indeed because the collision operator is purely local and the streaming operator requires shifting data just to nearest neighbors, we have seen no saturation of the parallelization to over 163 000 cores on $Blue Gene/Intrepid$ (Argonne) and on over 216 000 cores on $Jaguarpf$ (Cray XT-5 at Oak Ridge National Laboratory)---these being the maximum number of cores available to us currently. 

 \section{Poincare Recurrence for a class of initial conditions}
 
 We consider straight line vortices as the initial conditions for our simulations of the GP Eq. (22).  Unlike a classical vortex, a quantum vortex is a topological singularity with the wave function $\varphi = 0$ at the core singularity while the velocity $\bm{v}$, Eq. (4), diverges there.  Using Pade approximants \cite{berloff2004}, on the GP Eq. (22) one can determine an asymptotic steady state straight line vortex.  For winding number $n=1$, using cylindrical polar coordinates $(r,\phi, z$), such a vortex  that lies along the $z$-axis (and centered at the origin) is given by
\begin{equation}
\label{Pade_approximant}
\varphi (r) = e^{i \phi}\,\sqrt{\frac{11 a\, r^2 (12 + a \,r^2)}{g\,[384+ a\, r^2 (128 + 11 a \,r^2)]}}  ,
\end{equation}
 with $| \varphi | \to 1/\sqrt {g}$ as r $ \to \infty$, and $| \varphi | \sim r \sqrt {a/g} $ as r $\to 0$. 

The coherence length $\xi$, for this single line vortex, is typically defined as the distance from the core singularity to the position at which the absolute value of the wave function approaches its asymptotic value 
\begin{equation}
\label{coherence_length}
\xi = (\sqrt{a g \rho_0})^{-1} \sim a^{-1/2}.
\end{equation}
Quantitatively, the coherence length is defined from the solution of an isolated line vortex for a boundary value problem of the GP equation.  Here, we study quantum turbulence with many vortices interacting under periodic boundary conditions.   Alternatively, following Nore et. al. \cite{nore1997physfluid}, the coherence length can be defined from a \emph{linear} perturbation dispersion relation about a uniform density.   Qualitatively, in quantum turbulence with its many interacting quantum vortices, it is tempting to define a coherence length by replacing the background asymptotic density $\rho_0$ by the spatially averaged mean BEC density $< \rho_0>$, \cite{proment}.  However, in our simulations we rescale the initial wave function, Eq.(\ref{Pade_approximant}) so that our initial condition are far from a quasi-steady state solution to the GP Eq.(\ref{Gross_Pitaevskii_equation}) and so leads to turbulence more rapidly, especially on very large grids.  Under these conditions we feel that the standard idea of coherence length does not readily apply.

\begin{figure}[!h!t!b!p]
\begin{center}
\subfigure[\; winding no. n=1]{
\includegraphics[width=3.2in]{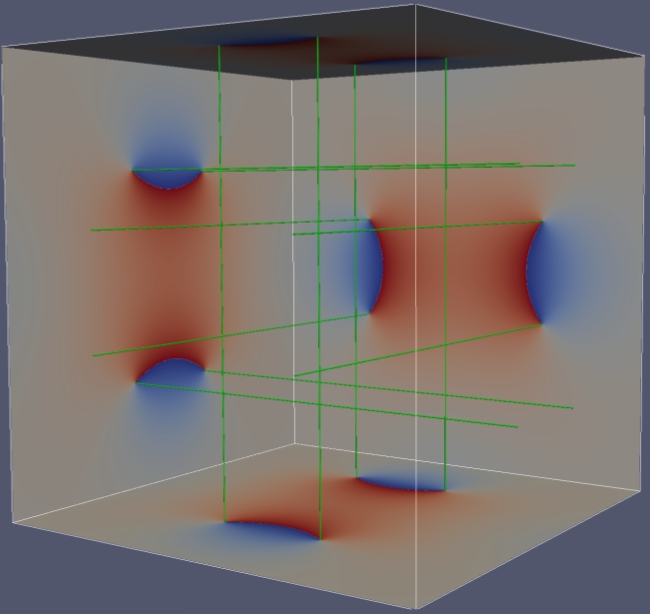}
}
\subfigure[\; winding no. n=2]{
\includegraphics[width=3.25in]{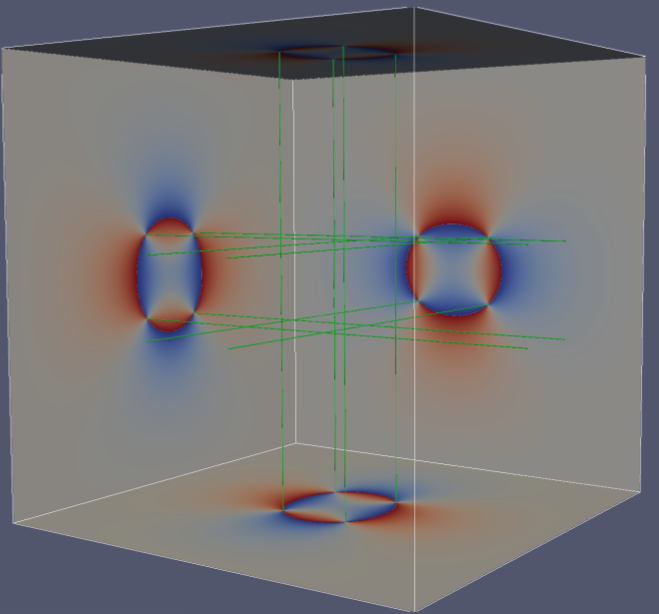}
}
\caption{\label{initial_isosurfaces} \footnotesize 
The initial isosurfaces of the $12$ vortex core singularities for (a)  winding number $n=1$, and (b) winding number $n=2$.  (a)  For winding number $n=1$, the core isosurface shown here is for $|\varphi|=0.13 |\varphi|_\text{max}$ since for lower isosurface values these initial cores would be too faint.  (b) For winding number $n=2$, the core singularities are shown for isosurface $|\varphi|=0.008 |\varphi|_\text{max}$  The initial location of the $12$ vortex cores for winding numbers $1$ and $2$ are different.  On the walls are shown the corresponding phases $\phi$, with $\phi = 0$ in blue and $\phi = 2 \pi$ in red.  The core singularity intercept with the walls acts like a branch point in the phase plane, with a branch cut joining the branch points. Grid $1200^3$.
 }
\end{center}
\end{figure}

For  winding number $n=1$ there is a phase change of $2 \pi$ in a closed circuit about the core singularity.  To enforce periodic boundary conditions we consider a set of four vortices parallel to the $z$-axis, with a similar set of four vortices with axes parallel to the $y$-axis and $x$-axis.  The total initial wave function is a product [4] of these twelve straight vortices and the initial quantum vortex core singularities are shown in  Fig. \ref{initial_isosurfaces} along with the phases on the boundaries of the lattice.  One immediately sees that at those points where the vortex isosurfaces intersect the plane walls there is a phase change from $\phi = 0$ (color blue) to $\phi = 2 \pi$ (color red).  Clearly there are $12$ such points.  Moreover, the sense of the BEC fluid rotation about the core singularity can be gleaned from the rotation sense of $\phi=0$ to $\phi= 2 \pi$.  Also shown in Fig. \ref{initial_isosurfaces} is a second set of initial conditions considered: a set of 12 doubly degenerate line vortices (essentially winding number of $n=2$ with $| \varphi | \sim r^{2} {a/g} $ as r $\to 0$) with phase change of $4 \pi$ around each core singularity.   The double degeneracy is evident from the phase information on the walls about these core singularity intersections and that this degeneracy is easily broken in the time evolution of the BEC as it is simply a confluence of two winding number $n=1$ singularities with energies significantly lower than that of the degenerate state.

\begin{figure*}[htbp!]
\begin{center}
\subfigure[\; $n=1$ at $t=6 000$]{
\includegraphics[width=2.78in]{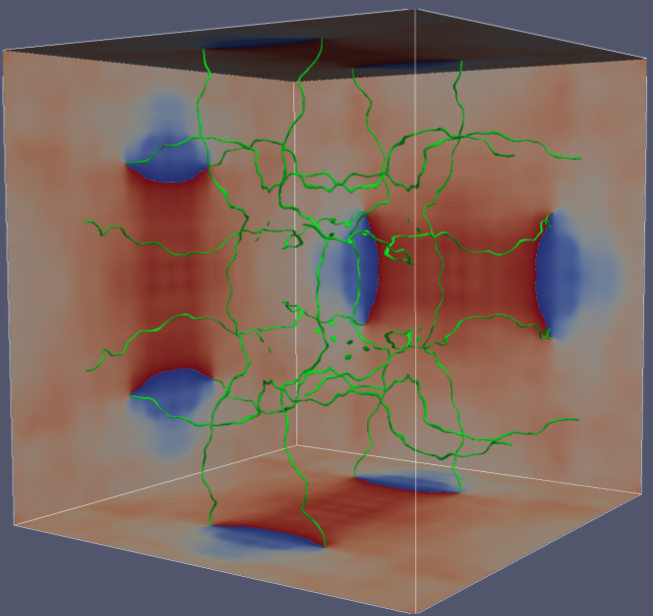}
}
\subfigure[\; $n=1$ at $t=6 000$ phase colored (blow-up)]{
\includegraphics[width=3.0in]{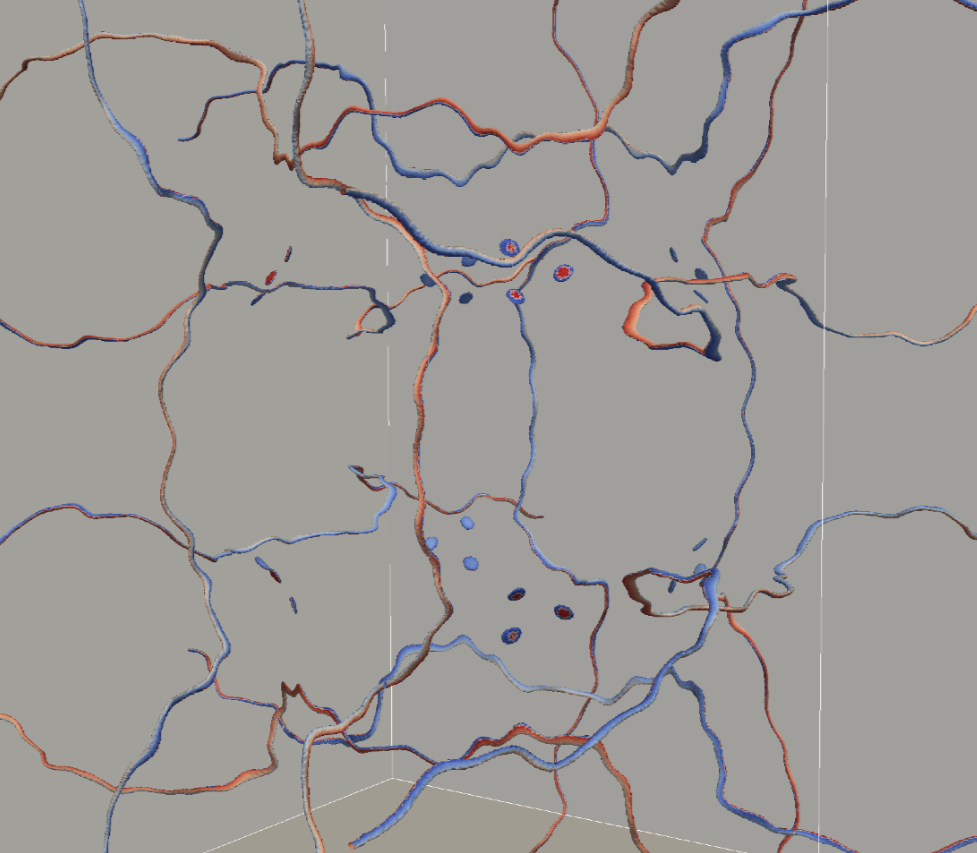}
}
\subfigure[\; $n=1$ at $t=25 000$]{
\includegraphics[width=2.9in]{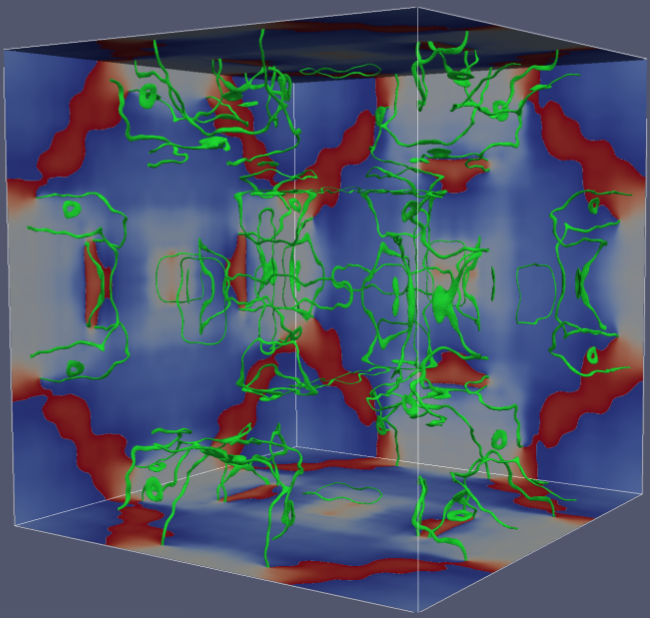}
}
\subfigure[\; $n=1$ at $t=25 000$ phase colored]{
\includegraphics[width=2.95in]{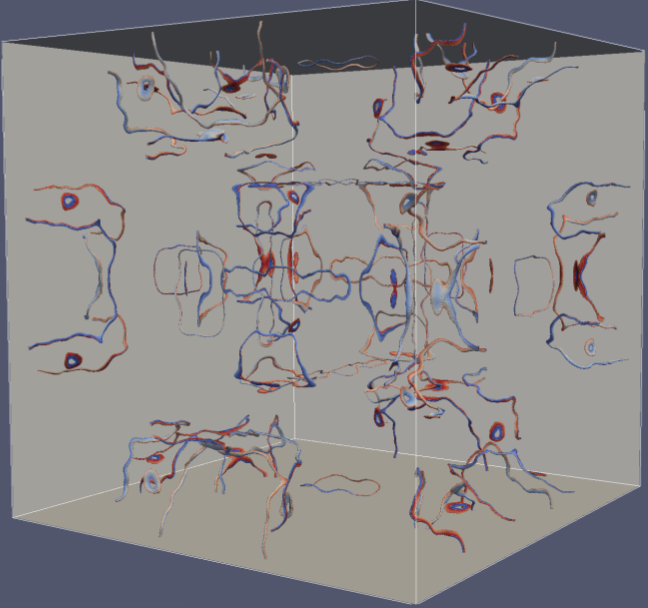}
}
\subfigure[\; $n=1$ at $t=99 000$]{
\includegraphics[width=2.85in]{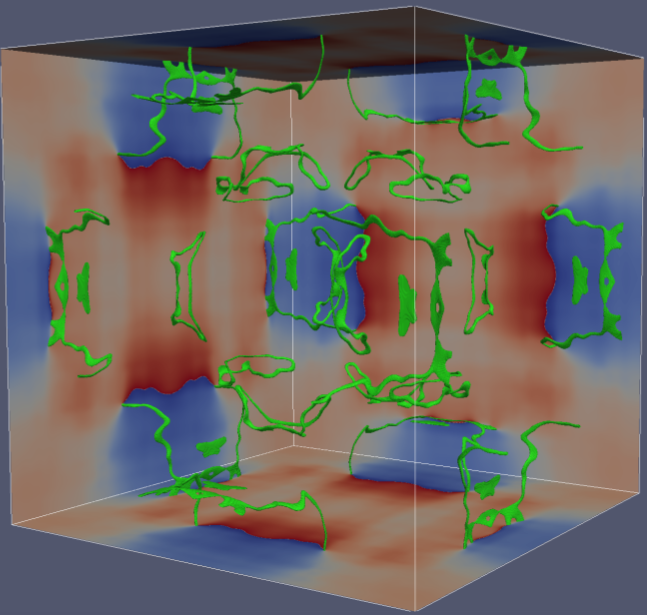}
}
\subfigure[\; $n=1$ at $t=99 000$ phase colored]{
\includegraphics[width=3.0in]{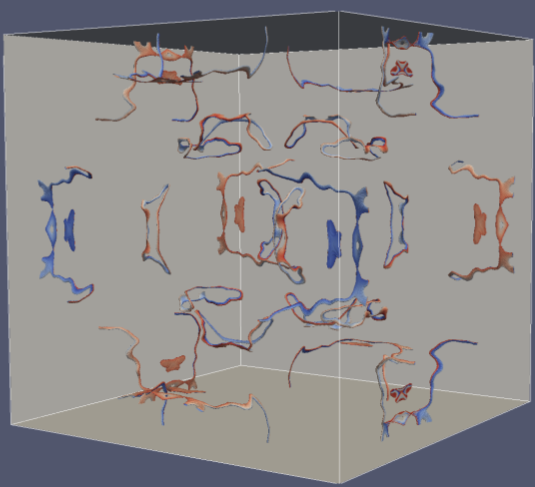}
}
\caption{\label{Poincare_small} \footnotesize 
The evolution of the isosurface core singularity (defined by  $|\varphi|=0.008 |\varphi|_\text{max}$) for winding number $n=1$ at times  (a)  $t=6000$, (b) also at  $t=6000$ but zoomed in perspective (same orientation) and now with phase information on the core itself; it is evident that there is a $2 \pi$ phase change around each vortex with vortex bending, reconnection, ring and blob formations.  Similarly (c) and (d) for time $t=25000$  and (e) and (f) for time
 $t=99000$.    Phase coding : $\phi = 0$ in blue, $\phi = 2 \pi$ in red.  Grid $1200^3$.
 }
\end{center}
\end{figure*}

\begin{figure*}[!h!t!b!p]
\begin{center}
\subfigure[\; $n=2$ at $t=6 000$]{
\includegraphics[width=3.0in]{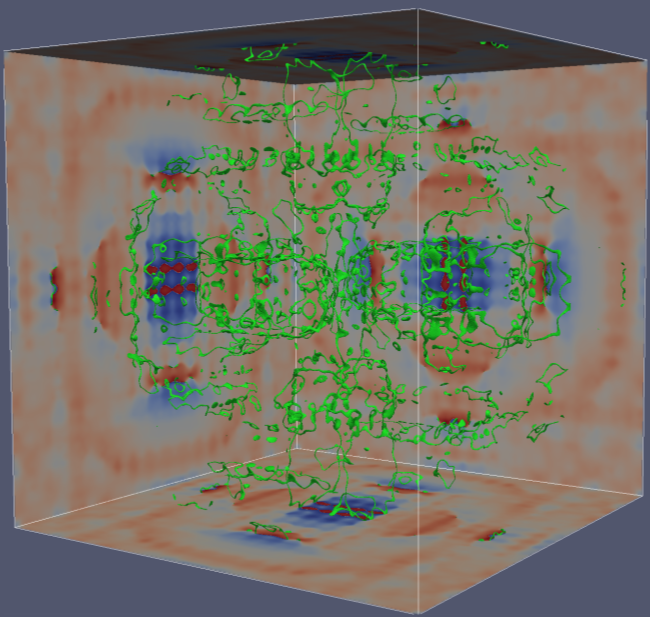}
}
\subfigure[\; $n=2$ at $t=6 000$ phase colored (blow-up)]{
\includegraphics[width=3.0in]{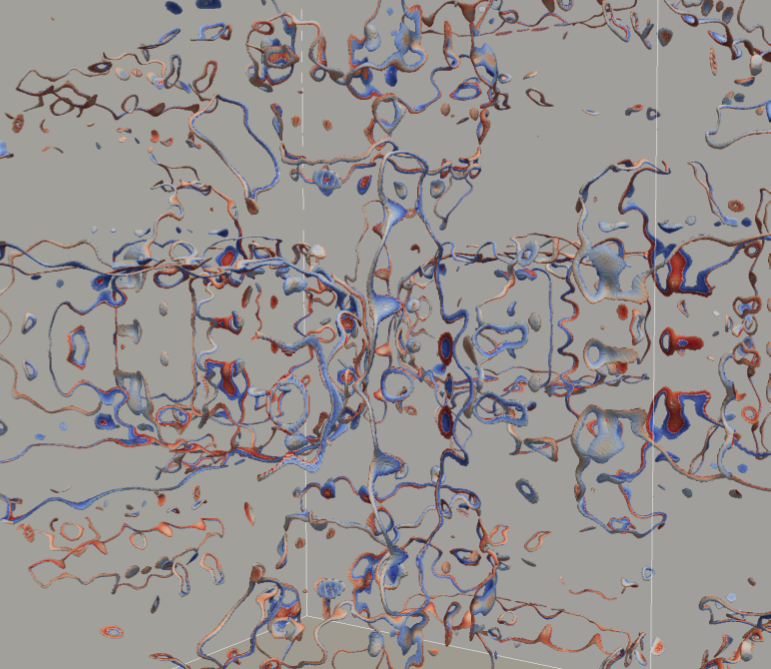}
}
\subfigure[\; $n=2$ at $t=25 000$]{
\includegraphics[width=3.00in]{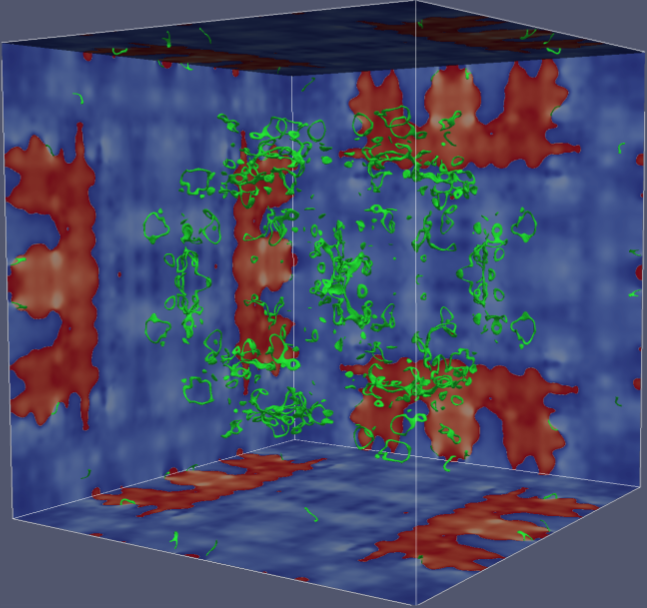}
}
\subfigure[\; $n=2$ at $t=25 000$ phase colored]{
\includegraphics[width=3.02in]{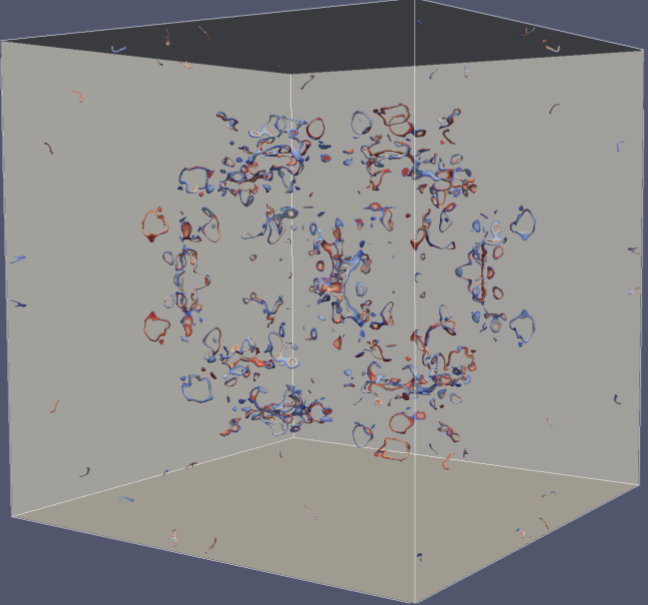}
}
\subfigure[\; $n=2$ at $t=99 000$]{
\includegraphics[width=3.00in]{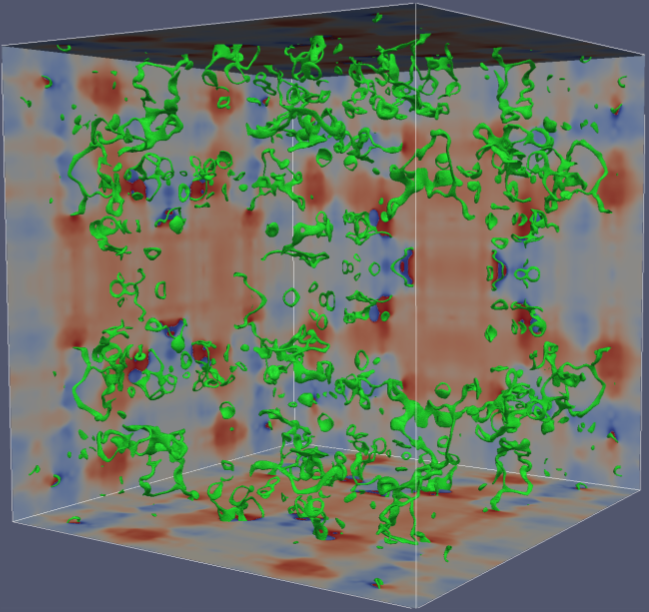}
}
\subfigure[\; $n=2$ at $t=99 000$ phase colored]{
\includegraphics[width=3.02in]{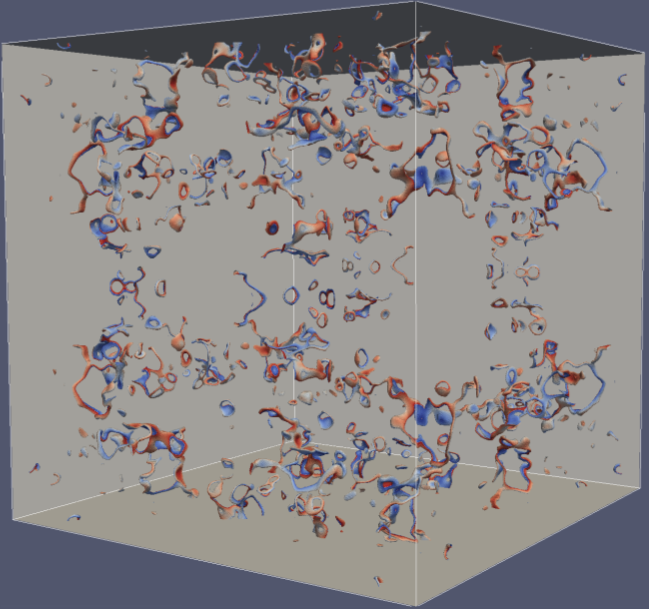}
}
\caption{\label{Poincare_n2} \footnotesize 
The corresponding isosurfaces of the core singularities for winding number $n=2$ at $t=6000$ (with zoomed-in perspective for phase-coded core singularity), $t = 25000$ and $t=99000$.   Phase coding : $\phi = 0$ in blue, $\phi = 2 \pi$ in red.  Grid $1200^3$
 }
\end{center}
\end{figure*}

We plot isosurfaces very close to the vortex core singularity.  The parameters and rescaling of the wave function are so chosen that $E_{int}(0) \ll E_{kin}(0), E_{qu}(0)$, with $\alpha(0) = \beta(0) = \varphi(0)/2$.  The early time evolution of the vortex core singularities are shown in Fig. \ref{Poincare_small} (for winding number $n=1$) and in Fig. \ref{Poincare_n2} (for winding number $n= 2$).  Here the phase information both on the walls  (with the green core singularity) and on the singularity core (with grey wall background) are plotted.  The $2 \pi$ phase change is evident when the singularity core intersects the walls [Figs.  \ref{Poincare_small}(a), (c) and (e)] while on close inspection one also sees the $2 \pi$ phase change along the singularity core itself, Fig.  \ref{Poincare_small}(b), (d) and (f).  Kevlin waves along the cores are clearly seen as are reconnections and vortex loops. 


As expected the  vortex core singularity structures for winding number $n=2$ (Fig. \ref{Poincare_n2}) are much more pronounced than for those core structures for winding number $n=1$ (Fig. \ref{Poincare_small}).  
 
\begin{figure}[!h!t!b!p]
\begin{center}
\subfigure[\; $n=1$]{
\includegraphics[width=3.4in]{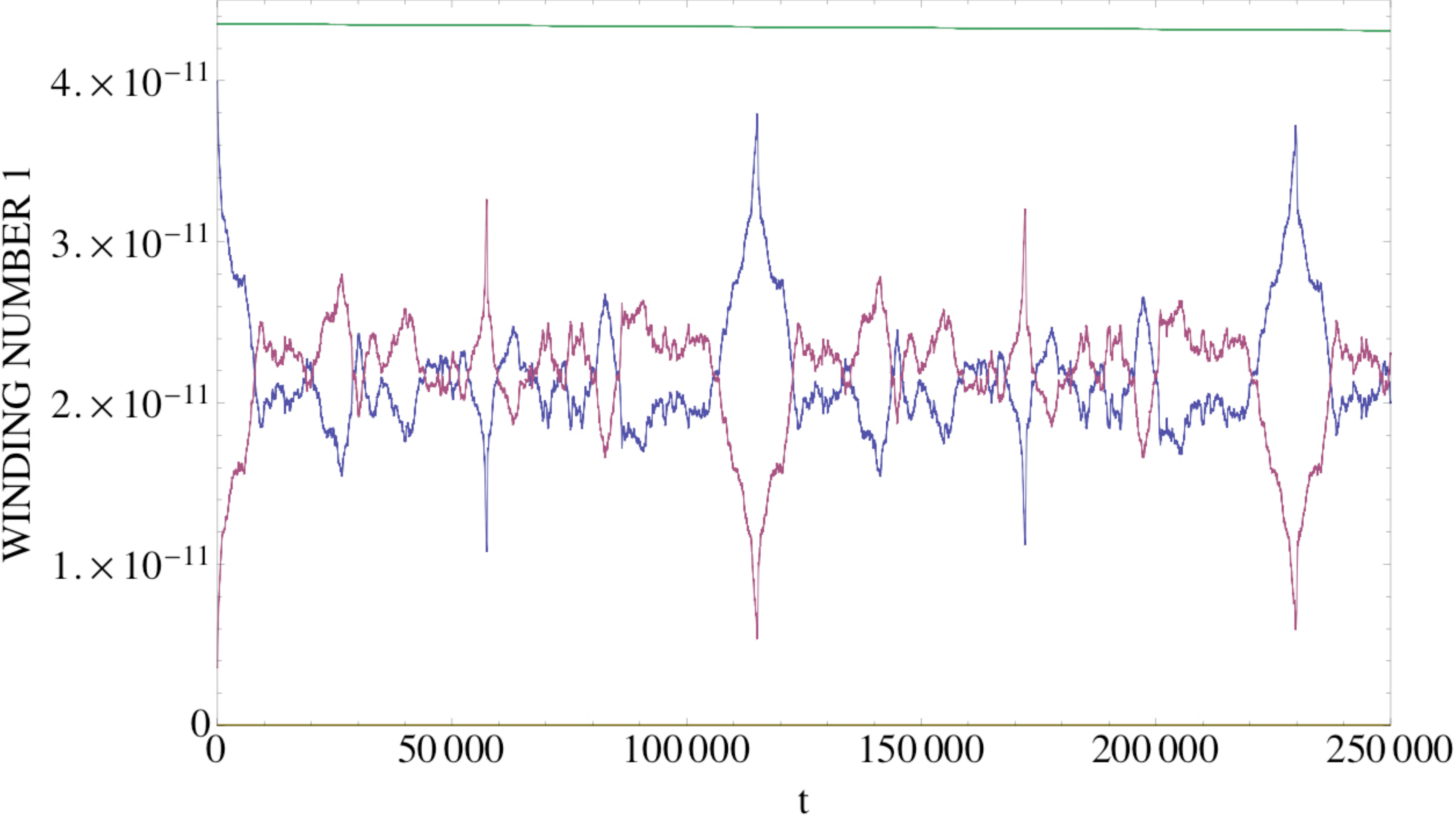}
}
\subfigure[\; $n=2$]{
\includegraphics[width=3.4in]{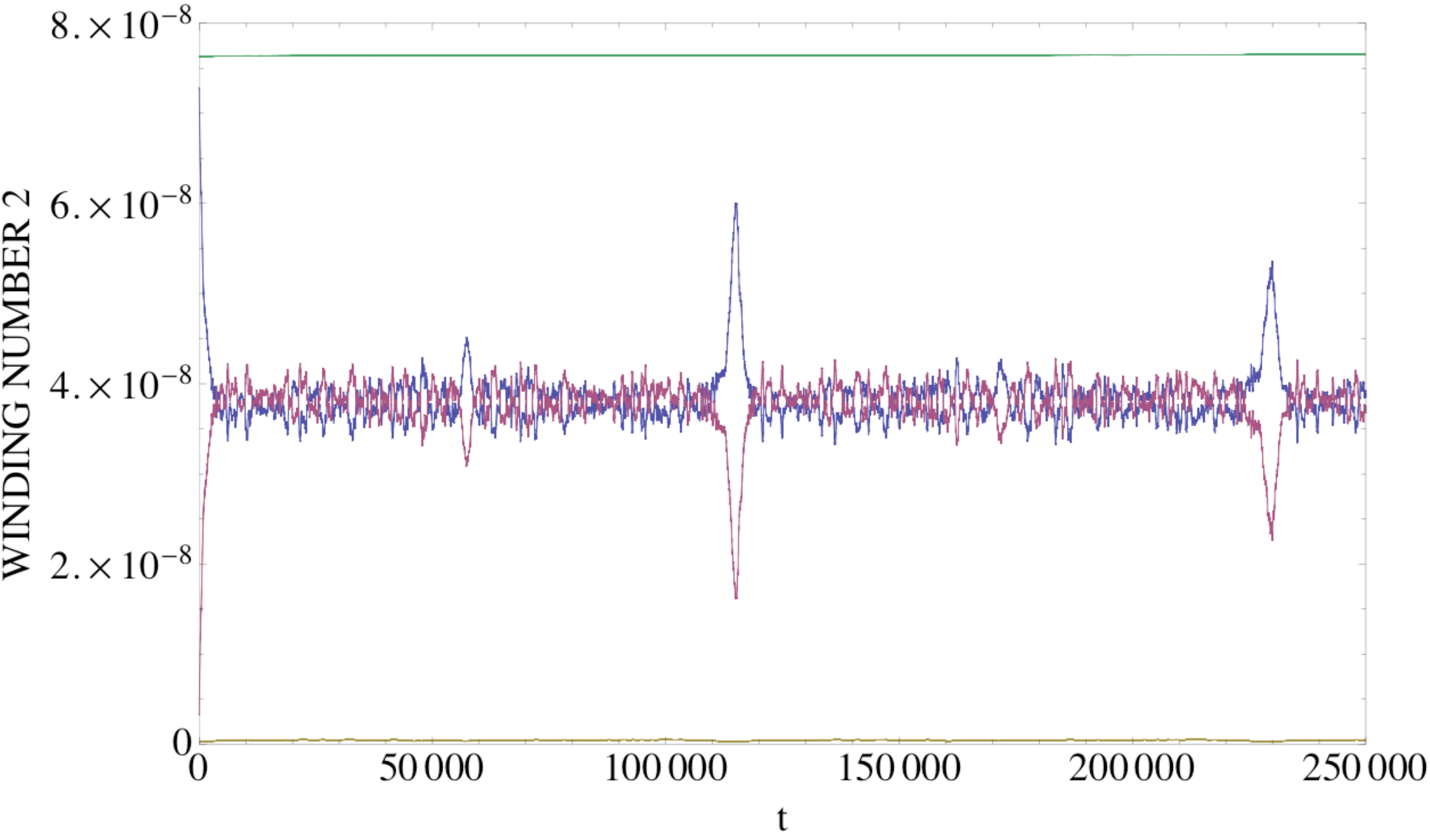}
}
\caption{\label{energy_evolution} \footnotesize 
The  time evolution of the $E_\text{kin}(t)$ (in blue)  and $E_\text{qu}(t)$ (in red) for $0\leq t \leq 250 000$ for (a) Winding Number $n=1$, and (b) Winding Number $n=2$. Throughout the run the unitary algorithm very well preserves the total energy conservation $E_\text{TOT} = const.$  The internal energy is negligible.  Note that $E_\text{kin}(0) \approx E_\text{kin}(t=115000) \approx E_\text{kin}(t=230000)  \approx\cdots$  for vortex cores with winding number $n=1$.  For vortex cores with winding number $n=2$, $E_\text{kin}(t=115000)$ and to a somewhat lesser extent $E_\text{kin}(t=230000)$ tend to $E_\text{kin}(t=0)$.
 }
 \end{center}
\end{figure}

The time evolution of the total energies, Eq. (3), is shown in Fig. \ref{energy_evolution} for vortex cores with (a) winding number $n=1$ and (b) with winding number $n=2$.  Note that the initial quantum vortices are so highly localized in space that the BEC density $\rho \approx$ const throughout the cubic lattice.  Thus $\nabla \sqrt \rho \approx 0$ throughout the lattice so that $E_{qu}(0) \ll E_{kin}(0)$ and $E_{comp}(0) \ll E_{incomp}(0)$.  

\begin{figure}[!h!t!b!p]
\begin{center}
\subfigure[\; $n=1$, $t = 115 000$]{
\includegraphics[width=3.0in]{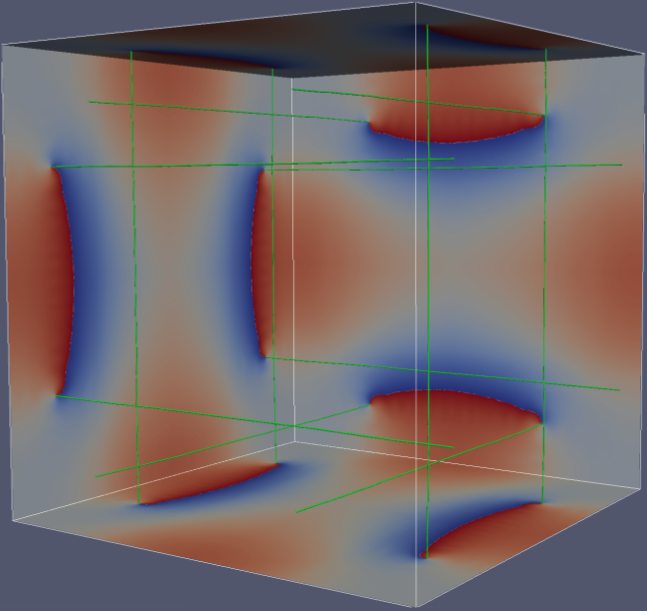}
}
\subfigure[\; $n=2$, $t = 115 000$]{
\includegraphics[width=3.0in]{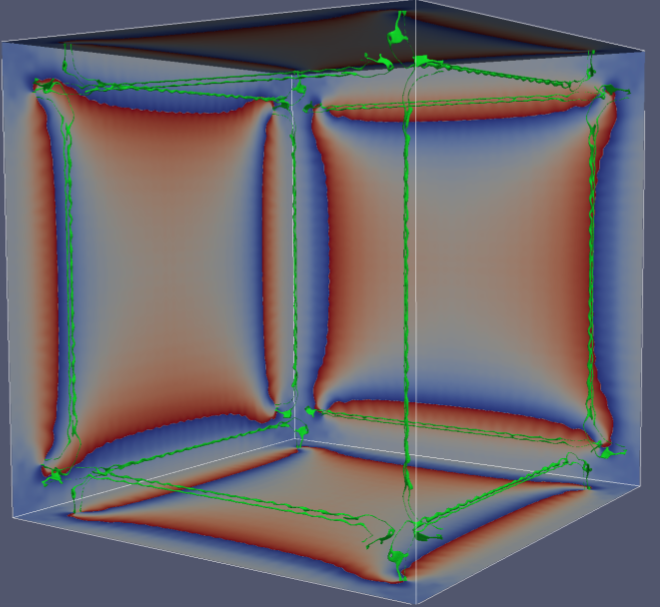}
}
\caption{\label{semi_Poincare_plot} \footnotesize 
The isosurface cores, with phase information on the walls, around the semi-Poincare period $t \simeq 115 000$ for (a)  winding number 1, and (b) winding number 2.  One sees a point inversion of the initial isosurface vortex coers.  For visibility, the vortex core isosurface for winding number 1 is shown at $|\varphi|=0.064 |\varphi|_\text{max}$, while those for winding number 2 at $|\varphi|=0.008 |\varphi|_\text{max}$.   Phase coding : $\phi = 0$ in blue, $\phi = 2 \pi$ in red.  Grid $1200^3$
 }
 \end{center}
\end{figure}

\begin{figure}[!h!t!b!p]
\begin{center}
\subfigure[\; $n=1$, $t = 230 000$]{
\includegraphics[width=3.02in]{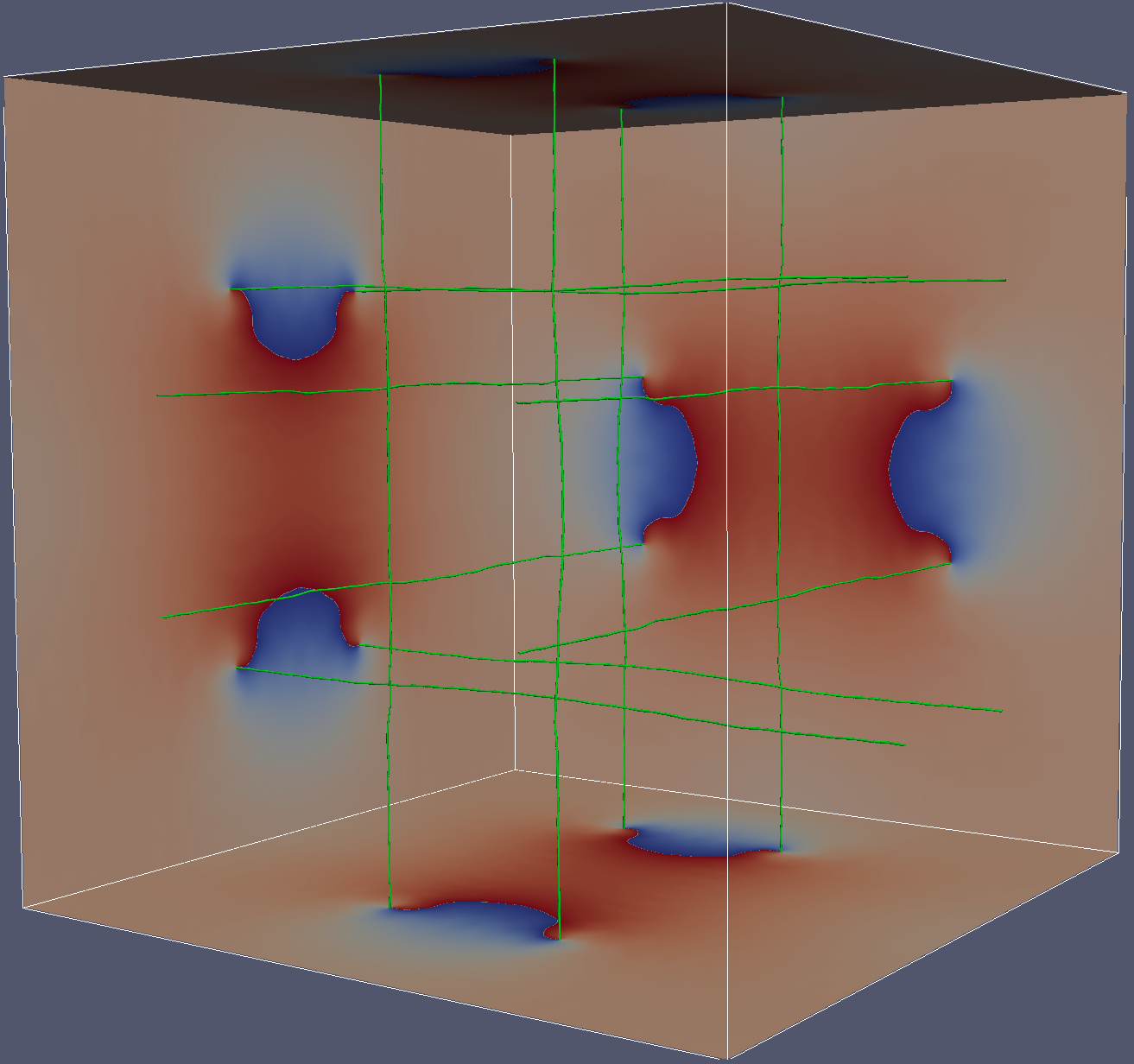}
}
\subfigure[\; $n=2$, $t = 230 000$]{
\includegraphics[width=3.0in]{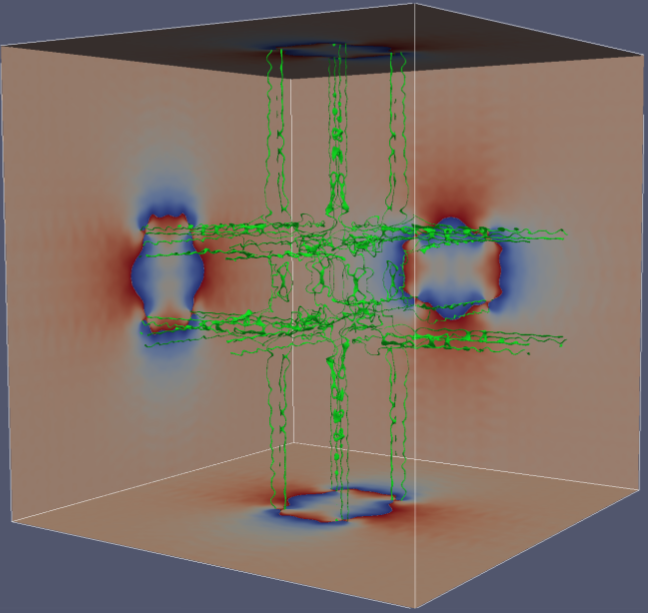}
}
\caption{\label{Poincare_plot} \footnotesize 
The isosurface cores around the Poincare period $t \simeq 230 000$ for (a) winding number 1, and (b) winding number 2.  The confluence degeneracy of winding number 2 singularities are now broken globally and there is considerable vortex loop generation locally.   Phase coding : $\phi = 0$ in blue, $\phi = 2 \pi$ in red.
 }
 \end{center}
\end{figure}

For quantum vortex cores with winding number $n=1$, from the time evolution of the kinetic and quantum energies one sees that $E_\text{kin}(0) \approx E_\text{kin}(t=115000) \approx E_\text{kin}(t=230000)  \approx\cdots$.  The situation is a bit more complex for cores with winding number $n=2$ in that the subsequent peaking in  $E_\text{kin}$ at $t =115000$ and $t=230000$ is somewhat reduced from its initial value.  This return of $E_\text{kin}(t)$ and $E_\text{qu}(t)$ close to their initial values is suggestive of a Poincare-recurrence, particularly as the GP equation is a Hamiltonian system.  To examine this we consider the corresponding vortex core isosurfaces for the different winding numbers at times $t = 115000$, Fig. \ref{semi_Poincare_plot} .  As expected, for winding number $n=2$, the confluence of the two singularities is broken with a slight split into two adjacent non-degenerate winding number $n=1$ singularities.
The vortex core singularity isosurfaces for winding number 1 at $t = 230000$ are shown in Fig. \ref{Poincare_plot} where it is clear that the Poincare recurrence time on a $1200^{3}$ grid is $T_{Poin} = 230000$.  On comparing Fig. \ref{initial_isosurfaces} with Fig. \ref{Poincare_plot} one does see some slight variations in the wall phase information from that at $t = 0$ as well as some small scale isosurface loop and vortex features now appearing for winding number 2.  

\begin{figure}[!h!t!b!p]
\begin{center}
\subfigure[\; initial image]{
\includegraphics[width=1.5in]{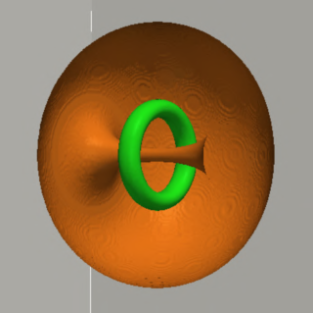}
}
\subfigure[\; iteration 80]{
\includegraphics[width=1.5in]{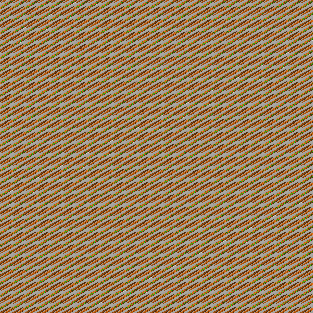}
}
\subfigure[\; point inversion symmetry, iteration $157=T_P/2$]{
\includegraphics[width=1.5in]{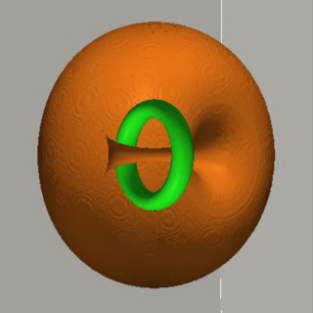}
}
\subfigure[\; Poincare recurrence at iteration $314 = T_P$]{
\includegraphics[width=1.5in]{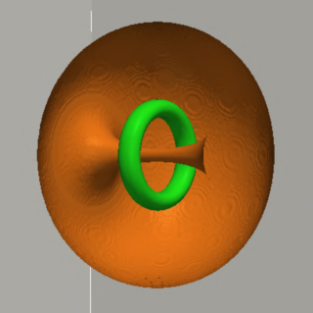}
}
\caption{\label{Skyrmion_inversion} \footnotesize 
The existence of point inversion symmetry for a given pixel resolution of an image under the Arnold Cat map $(x,y) \rightarrow (2x+y,x+y) mod \,1$.  This 2D map is invertible, area preserving, ergodic and mixing.   If the resolution of the skyrmion, (a), is $313 \times 313$, then at iteration $80$ one has the image (b).  At $T_P/2 = 157$, (c), one has the point inversion of the original skyrmion while at $T_P = 314$ one has Poincare recurrence.  However, for pixel resolution $315 \times 315$ there is no point inversion symmetry and $T_P = 120$.  (A skyrmion is a two-component topological defect with a vortex ring threaded by a quantum vortex line that then closes on the outer surface like an apple core).
 }
 \end{center}
\end{figure}

It is of some interest to note that the point inversion of the initial conditions at $T_{Poin}/2$  for the BEC wave function is similar to what can be seen in the discrete Arnold cat map for \emph{some}, but not all, pixel resolutions of the initial state - see Fig. \ref{Skyrmion_inversion}

\begin{figure*}[htbp!]
\begin{center}
\subfigure[\; $n=1$, $t = 0$ ,$48$ linear vortices, ]{
\includegraphics[width=2.8in]{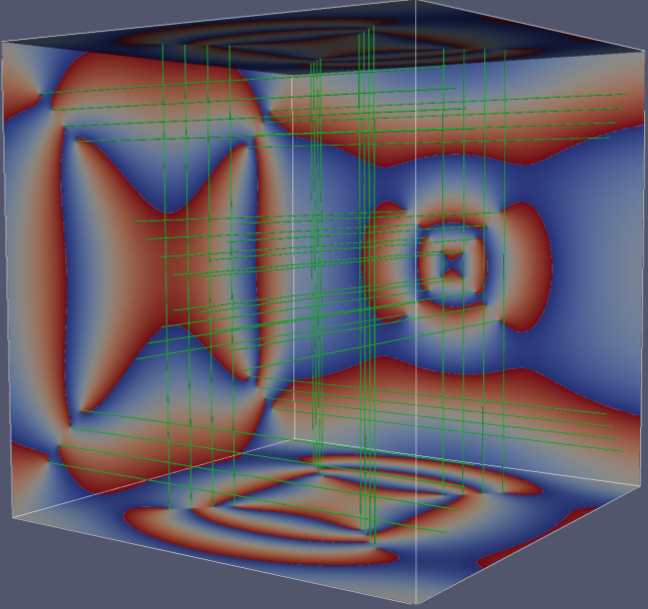}
}
\subfigure[\; $n=1$, $t = 84 000$]{
\includegraphics[width=2.80in]{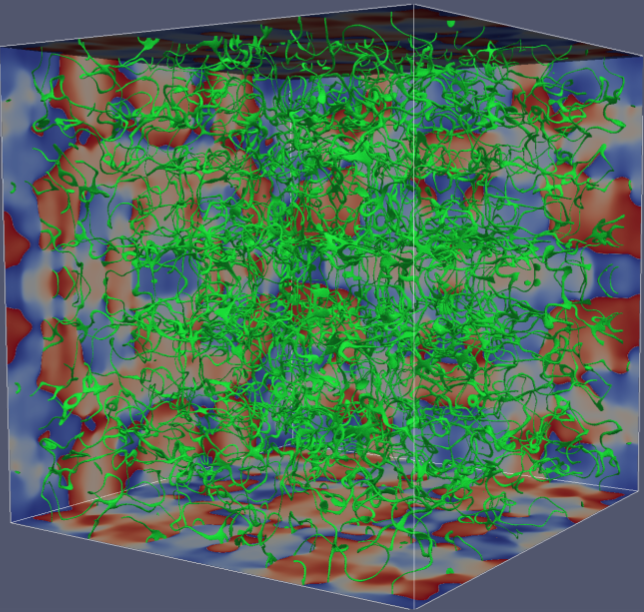}
}
\subfigure[\; $n=1$, $t = 115 000$]{
\includegraphics[width=2.80in]{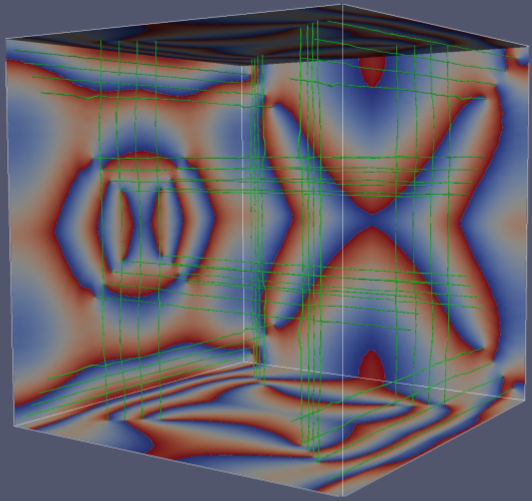}
}
\subfigure[\; $n=1$, $t = 230 000$]{
\includegraphics[width=2.80in]{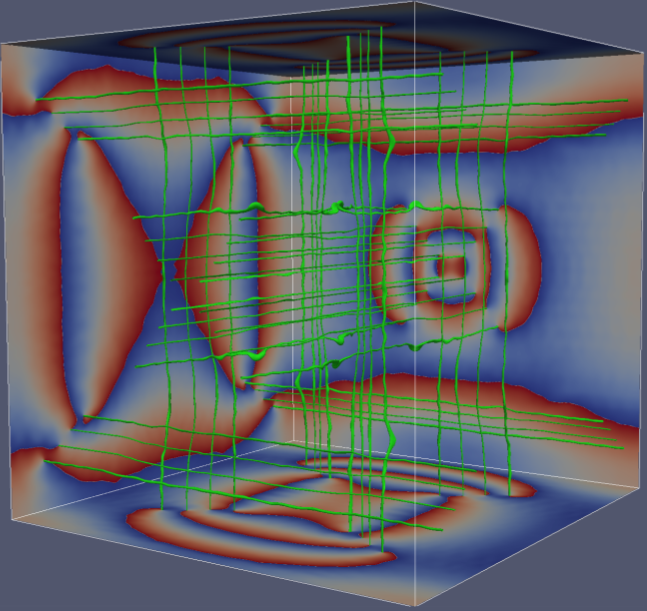}
}
\subfigure[\; $n=2$, $t = 0$ , $48$ linear vortices]{
\includegraphics[width=2.80in]{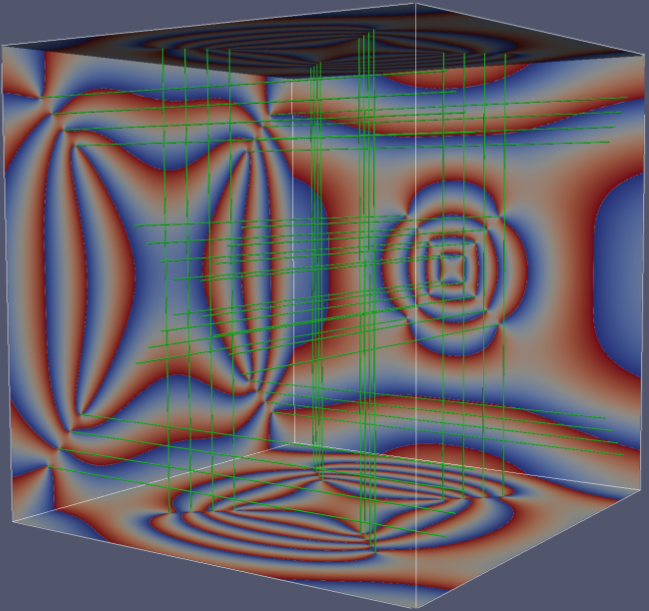}
}
\subfigure[\; $n=2$, $t = 230 000$]{
\includegraphics[width=2.80in]{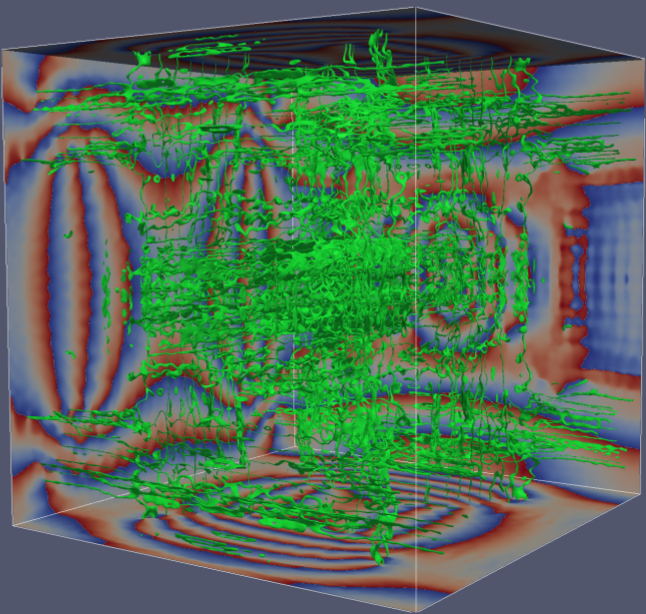}
}
\caption{\label{Poincare_plots_48} \footnotesize 
Evolution of quantum core singularities from an initial set of 48 straight line vortices with phase information on the walls.  (a) winding number $n=1$ at $t = 0$, (b) winding number $n=1$ at $t = 84000$. (c) winding number $n=1$ at $t = 230000 = T_{P}$ showing only small perturbative changes to the initial state given in (a).   (d)  winding number $n=1$ at $t = 115000 = T_{P}/2$.  The $2 \pi$ phase changes at the core singularity intersections at the walls is very evident.  (e)   winding number $n=2$ at $t=0$.  (f) The corresponding isosurface cores at $t = 230000 = T_{P}$ for winding number $n=2$.  The wall phase is a simple perturbative change from that at $t = 0$, (e) - but much small scale vortex structures due to the confluent degeneracy .   Phase coding : $\phi = 0$ - \emph{ blue}, $\phi = 2 \pi$ - \emph{red}.  Grid $1200^3$
 }
\end{center}
\end{figure*}

The robustness of the Poincare recurrence is further exhibited on simulating the evolution of 48 quantum vortex core singularities, Fig. \ref{Poincare_plots_48}.

Further, we find (see Table I) that the Poincare recurrence time, $T_{Poin}$, scales with diffusion ordering:  i.e., as the lattice grid scales from $L_{1} \rightarrow L_{2}$ the Poincare time rescales from $T_{Poin} \rightarrow (L_{2}/L_{1})^{2} T_{Poin}$.  It should be stressed that there is nothing per se in our quantum algorithm, Eq. (19), that enforces diffusion ordering at the GP level - to recover the GP equation it is critical to have diffusion ordering in our qubit unitary algorithm and this is achieved by specific choices of parameters in the algorithm itself.  Thus the Poincare recurrence follows physics ordering rather than the naively expected lattice ordering  $(L_{3}/L_{1})^{3}$ for 3D.

\begingroup
\begin{table}[thbp!]
\centering \caption{\footnotesize The scaling of the Poincare recurrence time, $T_{Poin}$ with the lattice grid $L$ assuming diffusion scaling and comparing this theoretical scaling with simulation results.  The scalings are normalized to a lattice grid of $512^3$.}   
\label{tab1} \smallskip
\begin{tabular}
{|l|l|l|l|} 
\hline 
{\sc Grid}
& 
{\sc Diffusion Scaling (theory) }
& 
{\sc Simulation }
\\
[0.25ex]
\hline 
$512^3$ 
& 
41775
& 
41775
 \\
[0.25ex]
\hline 
$960^3$ 
&
146850
&
147000
\\
[0.25ex]
\hline 
$1024^3$
& 
167100
& 
167600
 \\
[0.25ex]
\hline
 $1200^3$ 
& 
229477
& 
230000
\\
[0.25ex]
\hline
\end{tabular}
\end{table}
 \endgroup

\section{Intermittent Loss of the large-k Incompressible Kinetic Energy Spectral Cascade}
\begin{figure}
\begin{center}
\subfigure[\; $n=1$ , $t=0$]{
\includegraphics[width=2.5in]{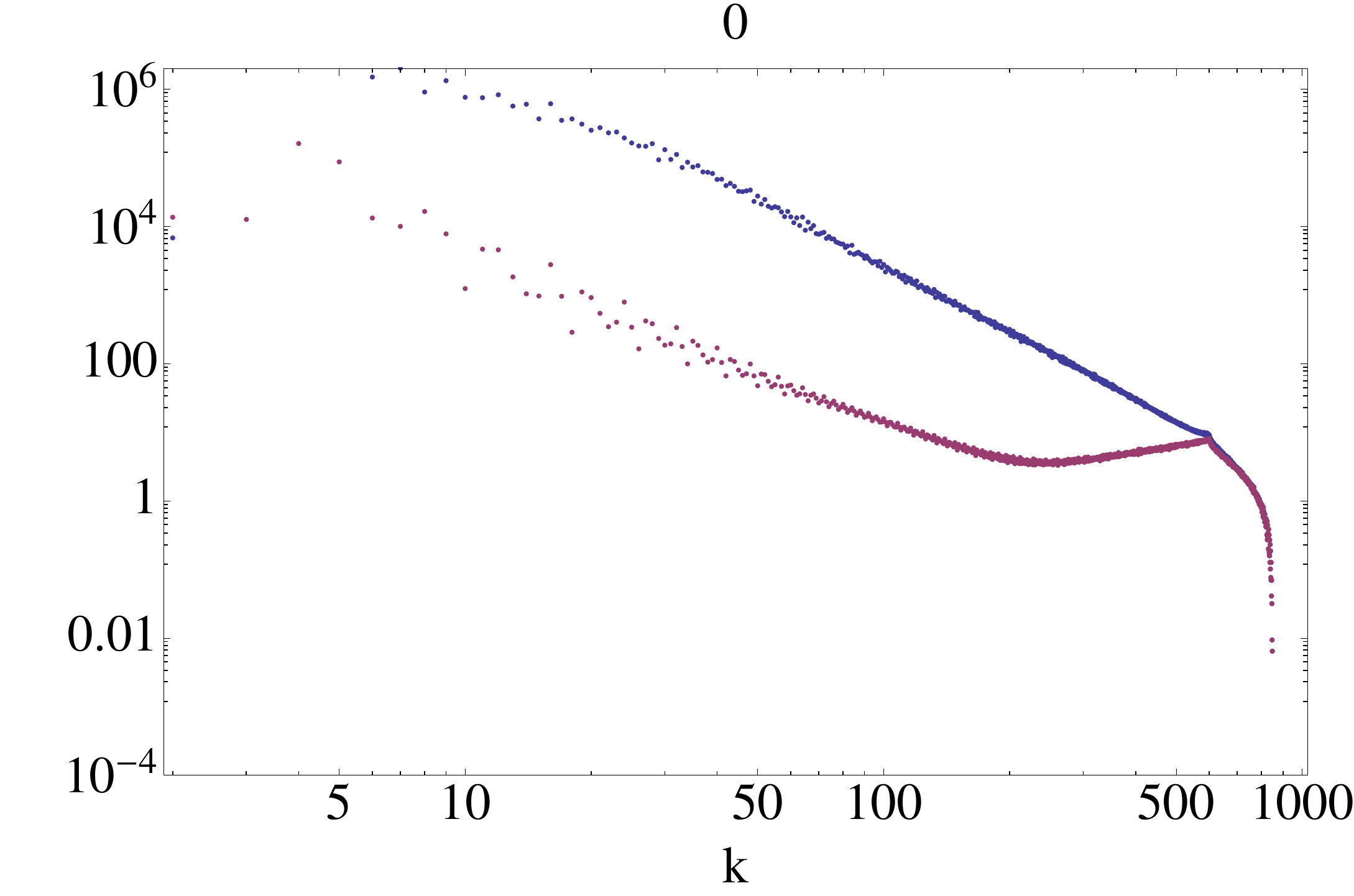}
}\\
\subfigure[\; $n=1$ , $t=20 000$]{
\includegraphics[width=2.5in]{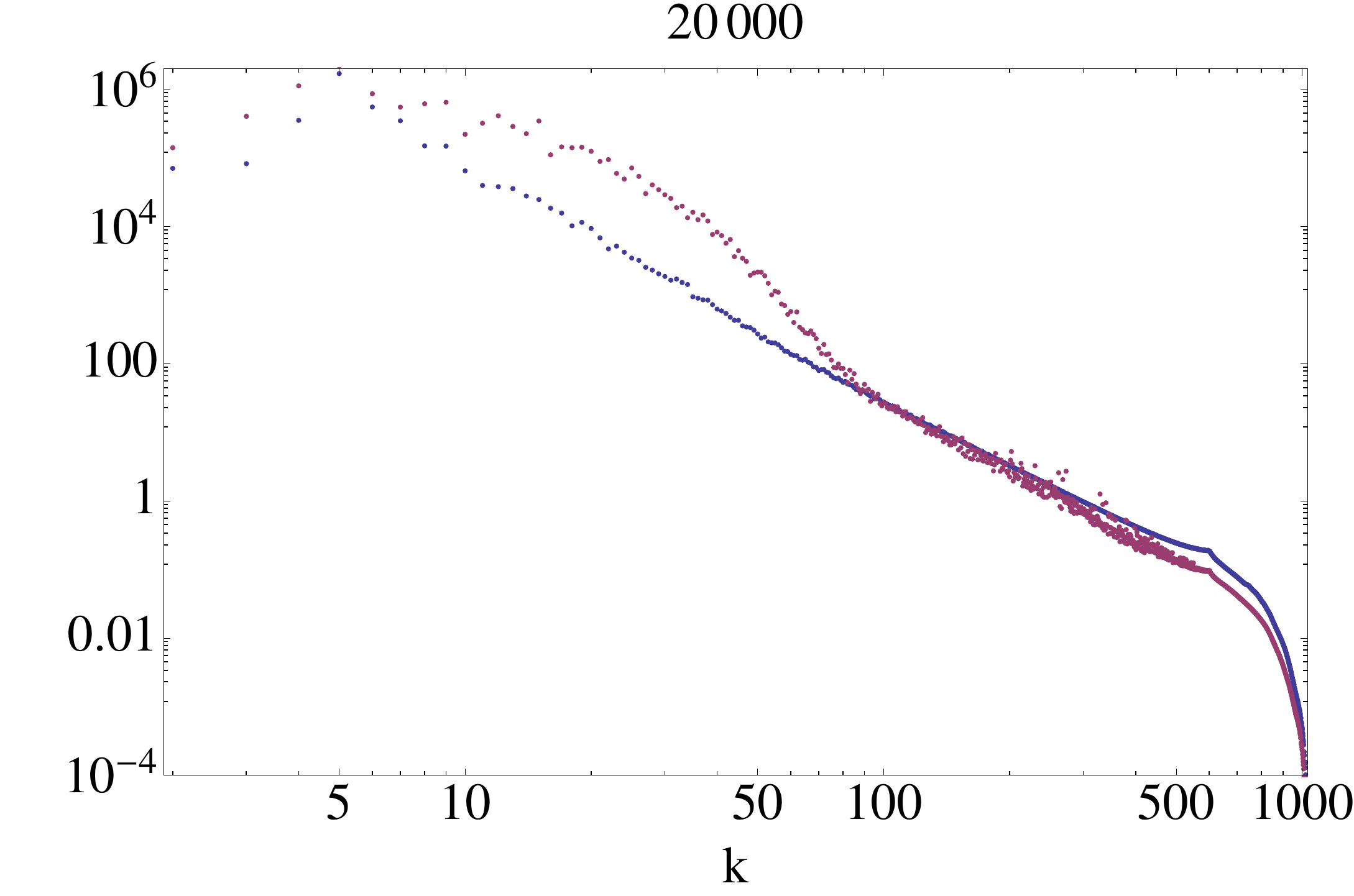}
}\\
\subfigure[\; $n=1$ , $t=75 000$]{
\includegraphics[width=2.5in]{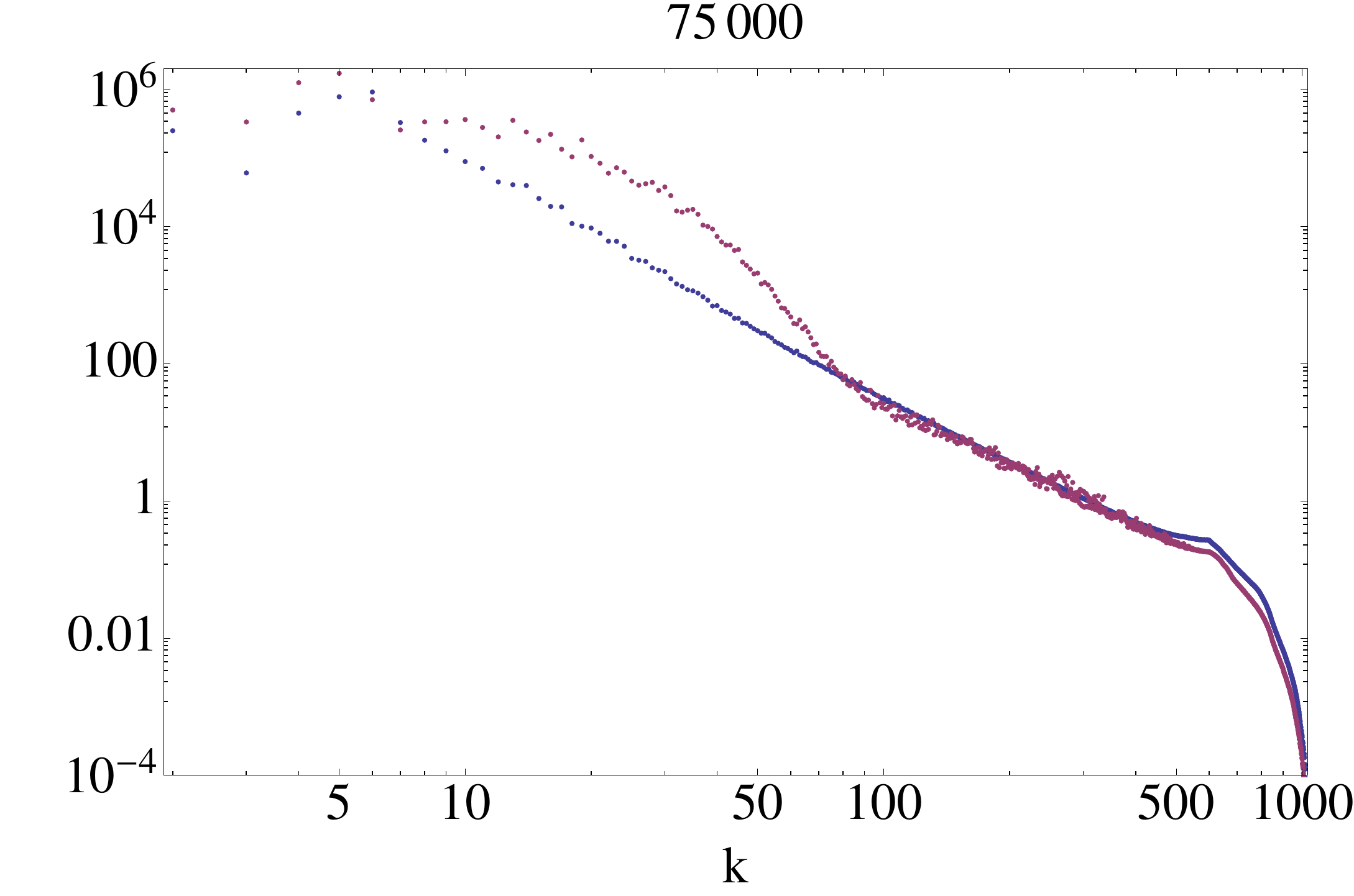}
}
\caption{\label{incomp_comp_spectra} \footnotesize 
Time snapshots of the incompressible (blue) and compressible (red) kinetic energy spectrum at times (a) t = 0, (b) t = 20 000, and (c) t = 75 000.   Initially nearly 99 \% of the kinetic energy is incompressible, while at t = 20 000 over 72\% of the kinetic energy is now compressible.  The spectra are quasi-steady state  for $8 000 < t < 81 000$.  The \emph{compressible} kinetic energy spectrum, $k^{-\alpha}$, exhibits 3 clear cascade regions: a classical Kolmogorov-like spectrum $\alpha \sim 5/3$ for $15<k<35$, a semi-classical cascade for $40<k<70$ and a quantum vortex spectrum with $\alpha \sim 3$ for $100<k<450$.  The \emph{incompressible} kinetic energy, on the other hand, exhibits a dual cascade spectrum $k^{-\alpha}$:  for small $k, \alpha \sim 3.7$ while for large $k,  \alpha = 3.0 $. Grid $1200^3$
 }
\end{center}
\end{figure}

An unexpected feature of our simulations is the intermittent loss of the incompressible kinetic energy spectral cascade for simple sets of initial line vortices with winding number $n=1$. This intermittency loss was first established from the time evolution of the incompressible kinetic energy spectrum and then verified by the topological behavior of the quantum vortex core singularities during this time interval.  To determine the incompressible kinetic energy spectra one first must decompose the density weighted velocity into its orthogonal compressible and incompressible components
%
\begin{equation}
\sqrt{\rho}  \bm{v} = [\sqrt{\rho}  \bm{v}]_{incomp}  +  [\sqrt{\rho}  \bm{v}]_{comp} ,
\qquad
\end{equation} 
with
\begin{equation}
\nabla \cdot [\sqrt{\rho}  \bm{v}]_{incomp} \equiv 0
\end{equation} 
%
and thus obtain \cite{nore1997physfluid} the compressible and incompressible kinetic energy spectra
\begin{eqnarray}
E_{kin}(t)
& \!\!=\!\!&
 E_{kin}^{incomp}(t) + E_{kin}^{comp}(t)
\\
& = &\int_{0}^{\infty} E_{kin}^{incomp}(k,t) dk + \int_{0}^{\infty} E_{kin}^{comp}(k,t) dk
\qquad
\end{eqnarray} 
For the initial $12$ straight line quantum vortices, over 98\% of the initial kinetic energy is incompressible, Fig. \ref{incomp_comp_spectra}(a)
 %

By $t > 8000$ the kinetic energy spectra have rapidly reached their quasi-steady state forms.  As can be seen in Figs. \ref{incomp_comp_spectra}(b) and (c), the \emph{compressible} kinetic energy spectrum exhibits 3 cascade regions while the \emph{incompressible} kinetic energy spectrum exhibits basically two slopes.  It should be pointed out that we had an error in our FFTs in \cite{yepez2009prl} and the spectra exhibited is basically the \emph{compressible} spectrum---see also \cite{yepez2010}.  To examine the Poincare recurrence of initial conditions and the intermittent loss of the vortex spectrum, we consider grids of $1200^{3}$ -- for example, for grids of $3072^3$, the first Poincare recurrence would occur around t = 720 000 and require much wallclock time and cores with little new physics insights.  For $t > 8000$ the \emph{compressible} kinetic energy spectrum $k^{-\alpha}$ has 3 distinct cascade regions:  a classical Kolmogorov-like energy cascade for $15 < k < 35$ with exponent $\alpha \sim {5/3}$, a steep semi-classical cascade for $40 < k < 70$ and a quite long quantum vortex spectrum for $100 < k < 450$ with exponent $\alpha \approx 3$.  On the otherhand, the \emph{incompressible} kinetic energy spectrum exhibits a dual cascade over nearly all wave number range with exponents $\gamma \approx 3.7$ and $\gamma = 3.0$.  Since for $t>8000$, $E_{kin}^{comp}(t) > 3.5 E_{kin}^{incomp}(t)$ and from Fig. \ref{incomp_comp_spectra} one can readily see that the \emph{total} kinetic energy spectrum will also exhibit these 3 cascades will little change to the corresponding exponents. 

\begin{figure*}[htbp!]
\begin{center}
\subfigure[\; $n=1$ , $t=81 400$]{
\includegraphics[width=2.15in]{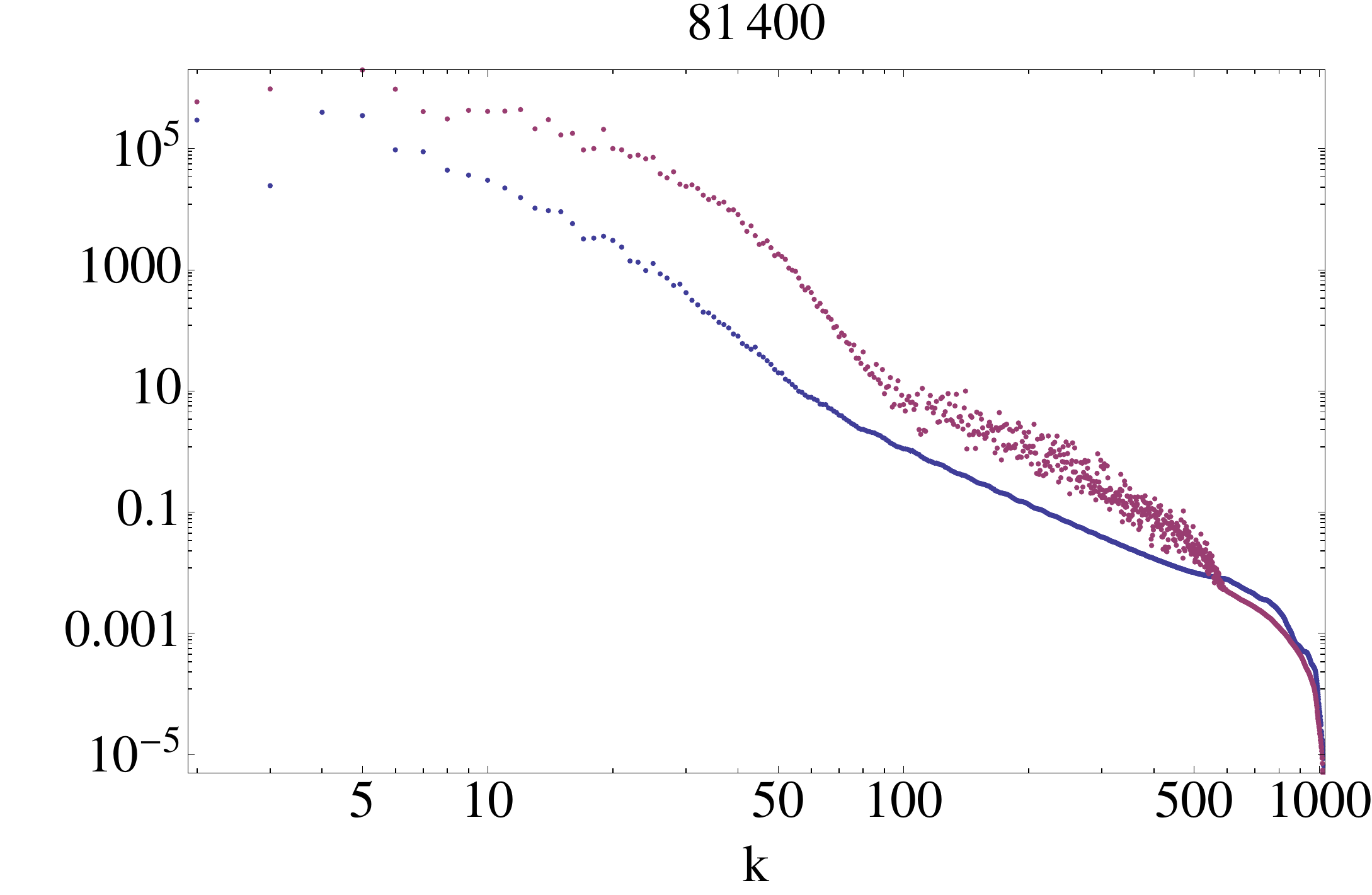}
}
\subfigure[\; $n=1$ , $t=82 200$]{
\includegraphics[width=2.15in]{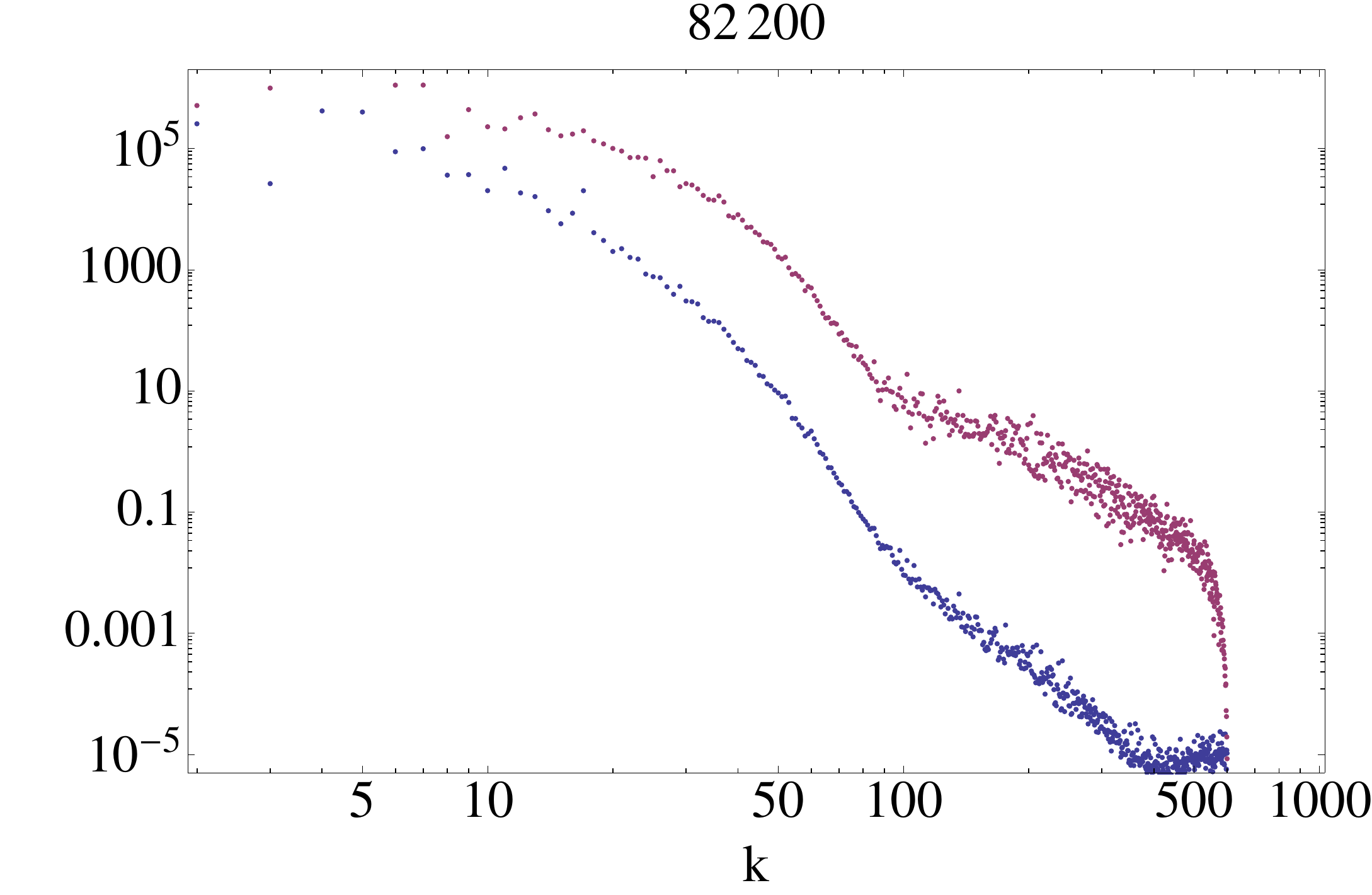}
}
\subfigure[\; $n=1$ , $t=83 400$]{
\includegraphics[width=2.15in]{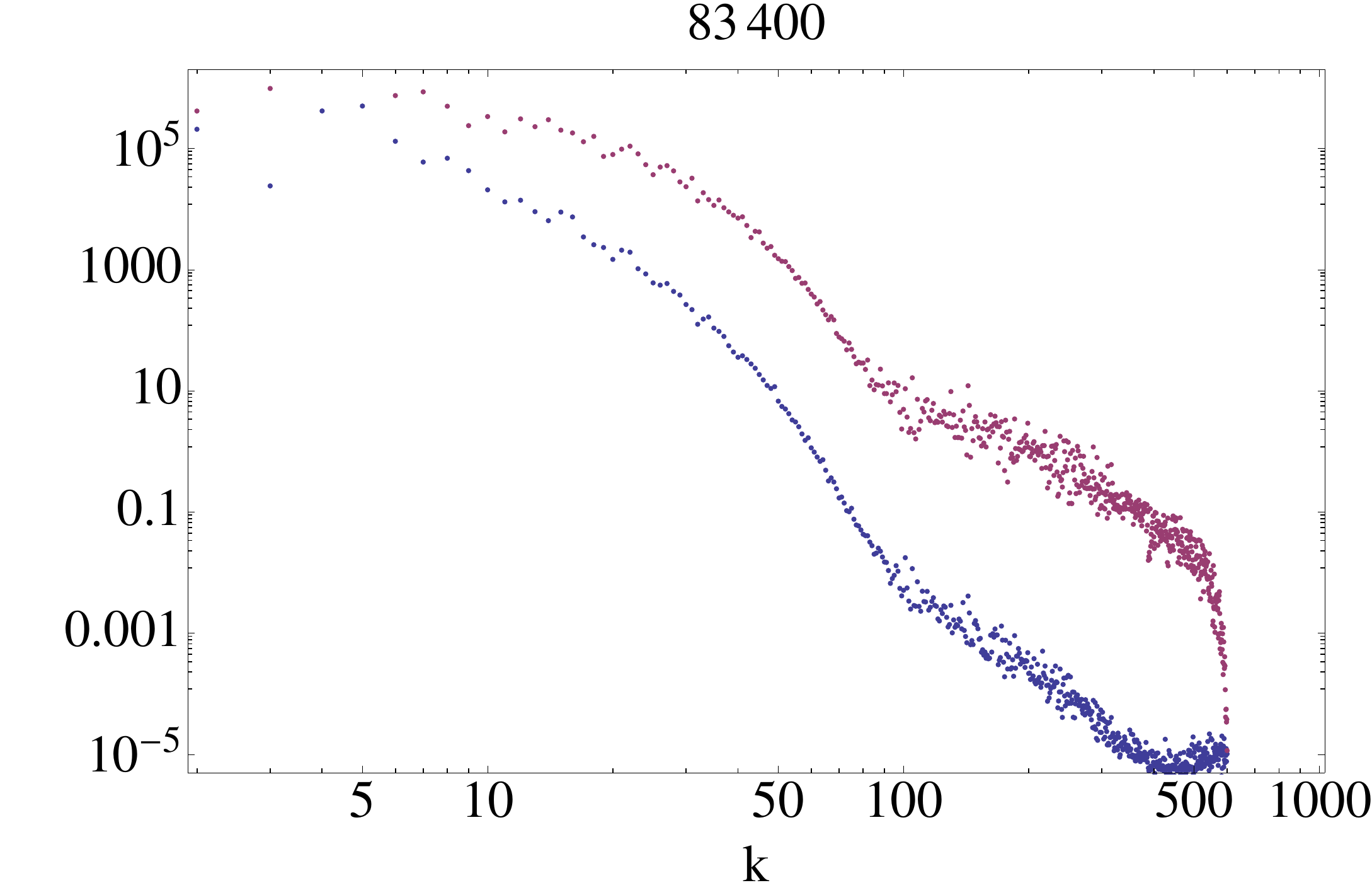}
}
\subfigure[\; $n=1$ , $t=84 000$]{
\includegraphics[width=2.15in]{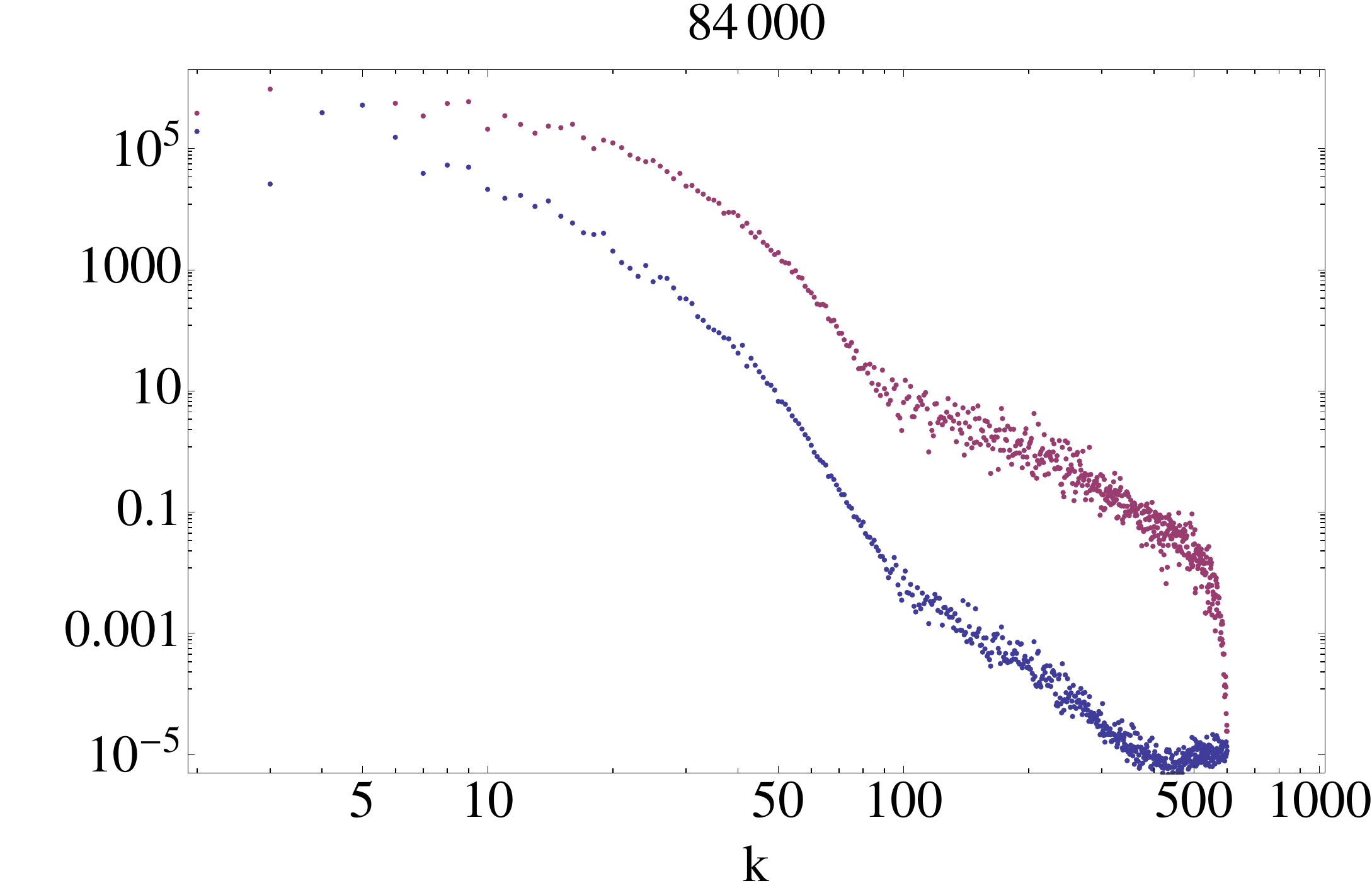}
}
\subfigure[\; $n=1$ , $t=84 200$]{
\includegraphics[width=2.15in]{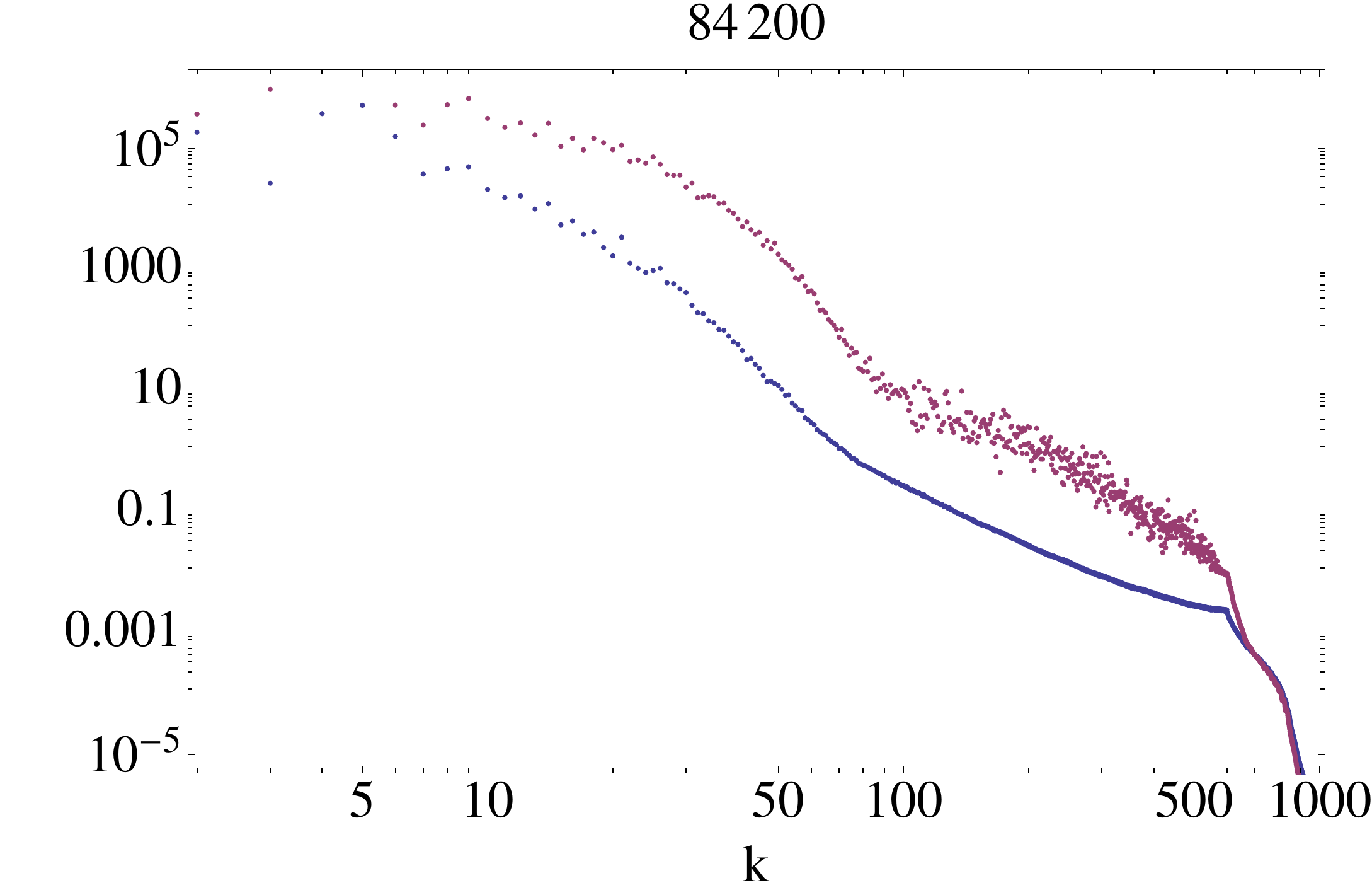}
}
\subfigure[\; $n=1$ , $t=85 000$]{
\includegraphics[width=2.15in]{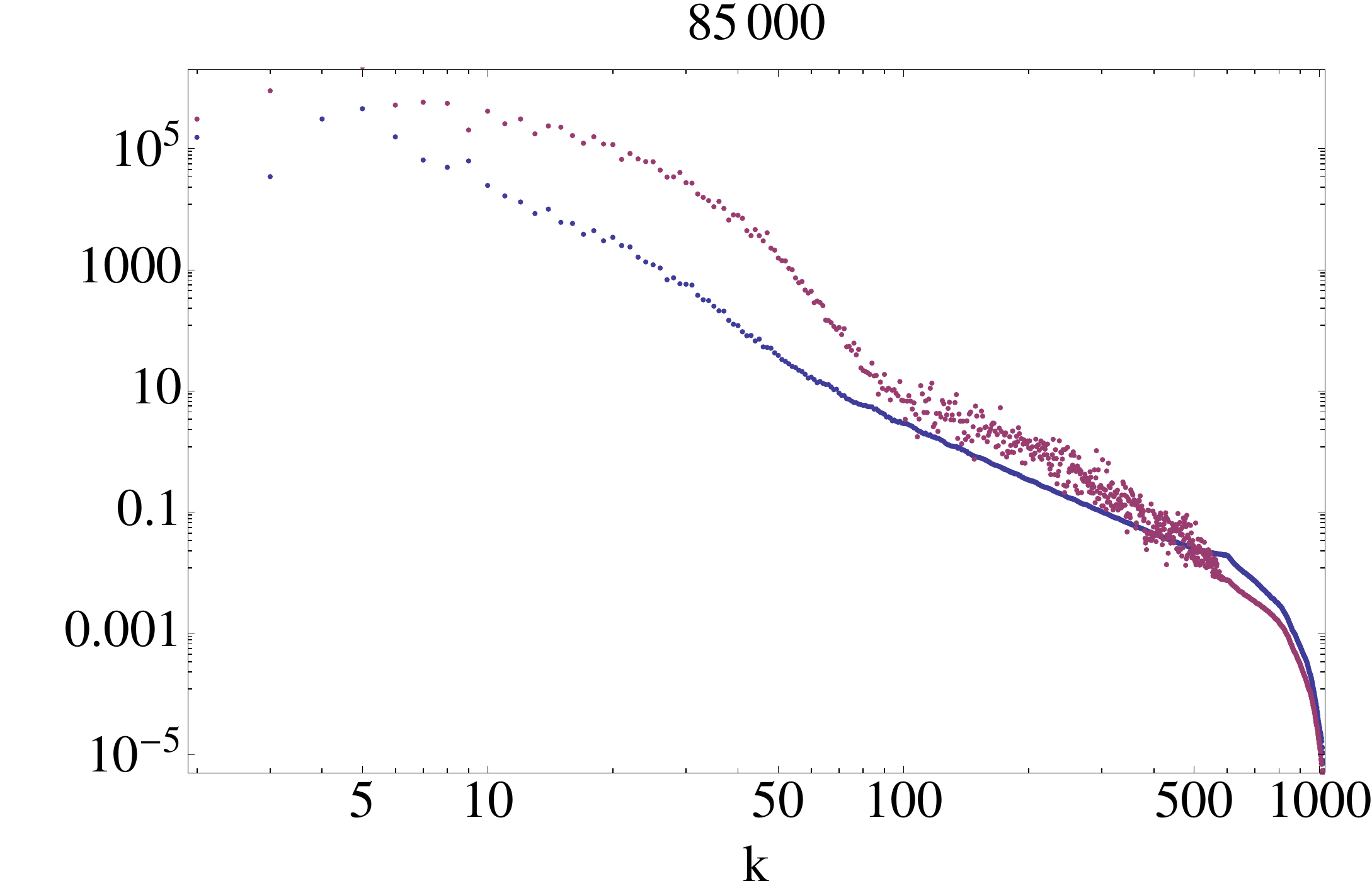}
}
\caption{\label{loss_incomp_spectrum} \footnotesize 
6 snapshots of the incompressible (blue) and compressible (red) kinetic energy spectrum at times (a) t = 81 400, (b) t = 82 200, (c) t = 83 400, (d) t = 84 000, (e) t = 84 200, and (f) t = 85 000.  In (b)-(d) there is a sharp drop in the incompressible energy spectrum for wave numbers $k>100$, except for a very brief transient recovery around $t \sim 83 000$.  There is also a sharp cutoff in the compressible spectrum for $k>500$. This signals the loss of the quantum vortex spectrum in this time interval.  Around these intermittencies, the incompressible kinetic energy spectrum also exhibits a triple cascade $k^{-\alpha}$ with $\alpha \sim 3.7$ for small $k$, an $\alpha \sim 6$ for the intermediate cascade, and $\alpha \sim 3.0$ for the large-$k$ quantum vortex spectrum.  During the intermittency, the large-$k$ exponent increases to a noisy $\alpha \sim 5.2$ as well as a steeped semi-classical intermediate exponent.  Grid $1200^3$
 }
\end{center}
\end{figure*}

For $81 400 < t < 84 300$, the kinetic energy spectra intermittently take the forms shown in Fig. \ref{loss_incomp_spectrum}.  The incompressible spectrum now takes on a quasi $3$ cascade spectrum with a brief loss in the $k^{-3}$ spectral region.

\begin{figure*}[!h!t!b!p]
\begin{center}
\subfigure[\; $n=1$ , $t=78 000$]{
\includegraphics[width=3.0in]{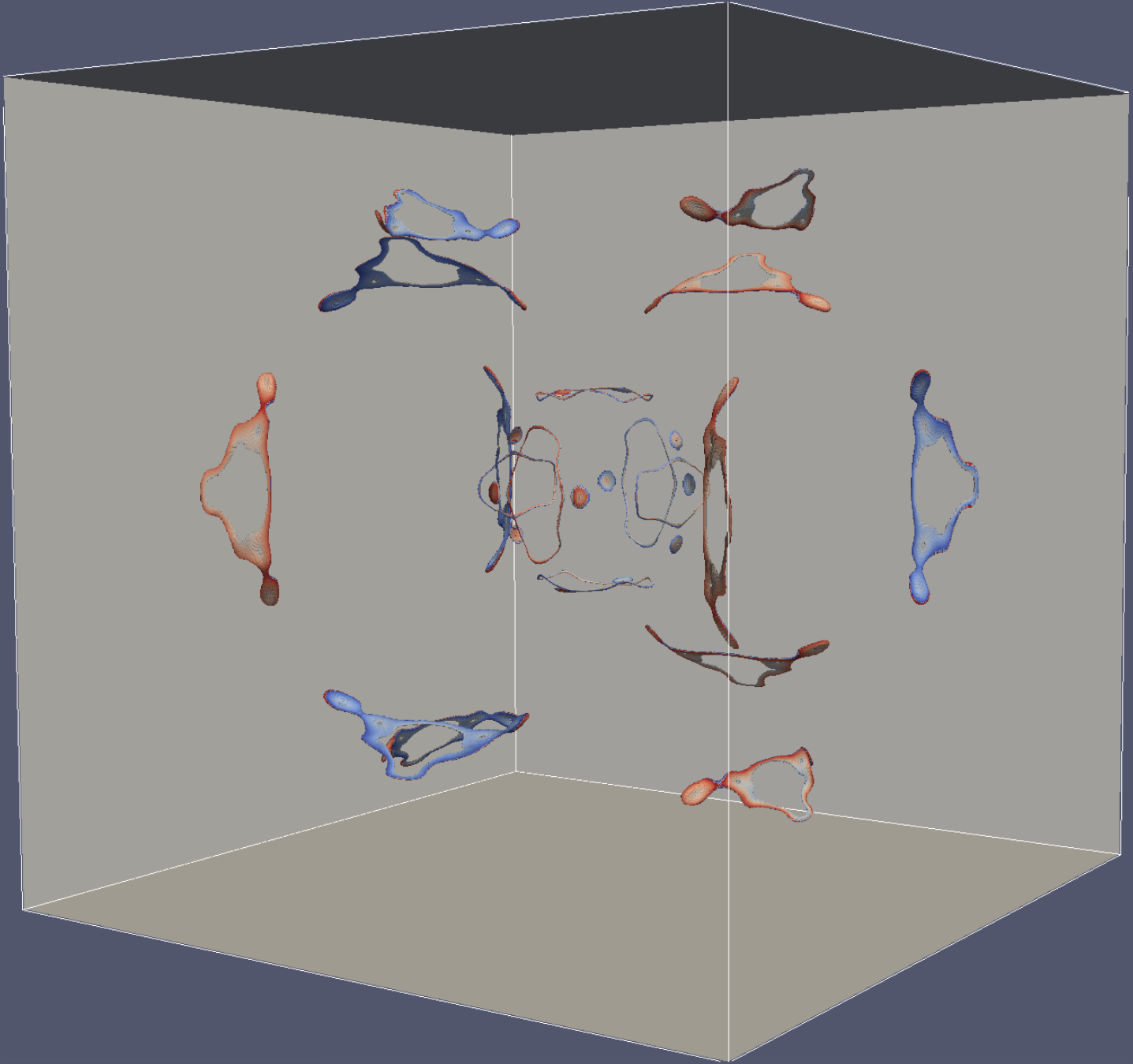}
}
\subfigure[\; $n=1$ , $t=81 000$]{
\includegraphics[width=3.0in]{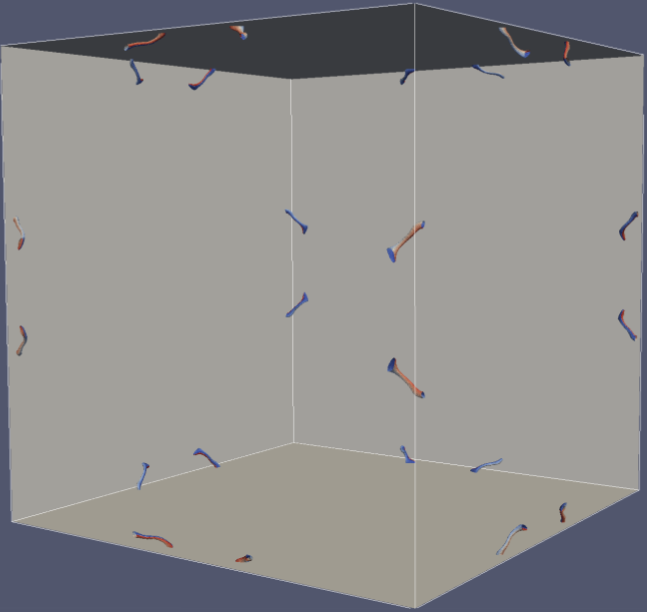}
}
\subfigure[\; $n=1$ , $t=82 000$]{
\includegraphics[width=3.0in]{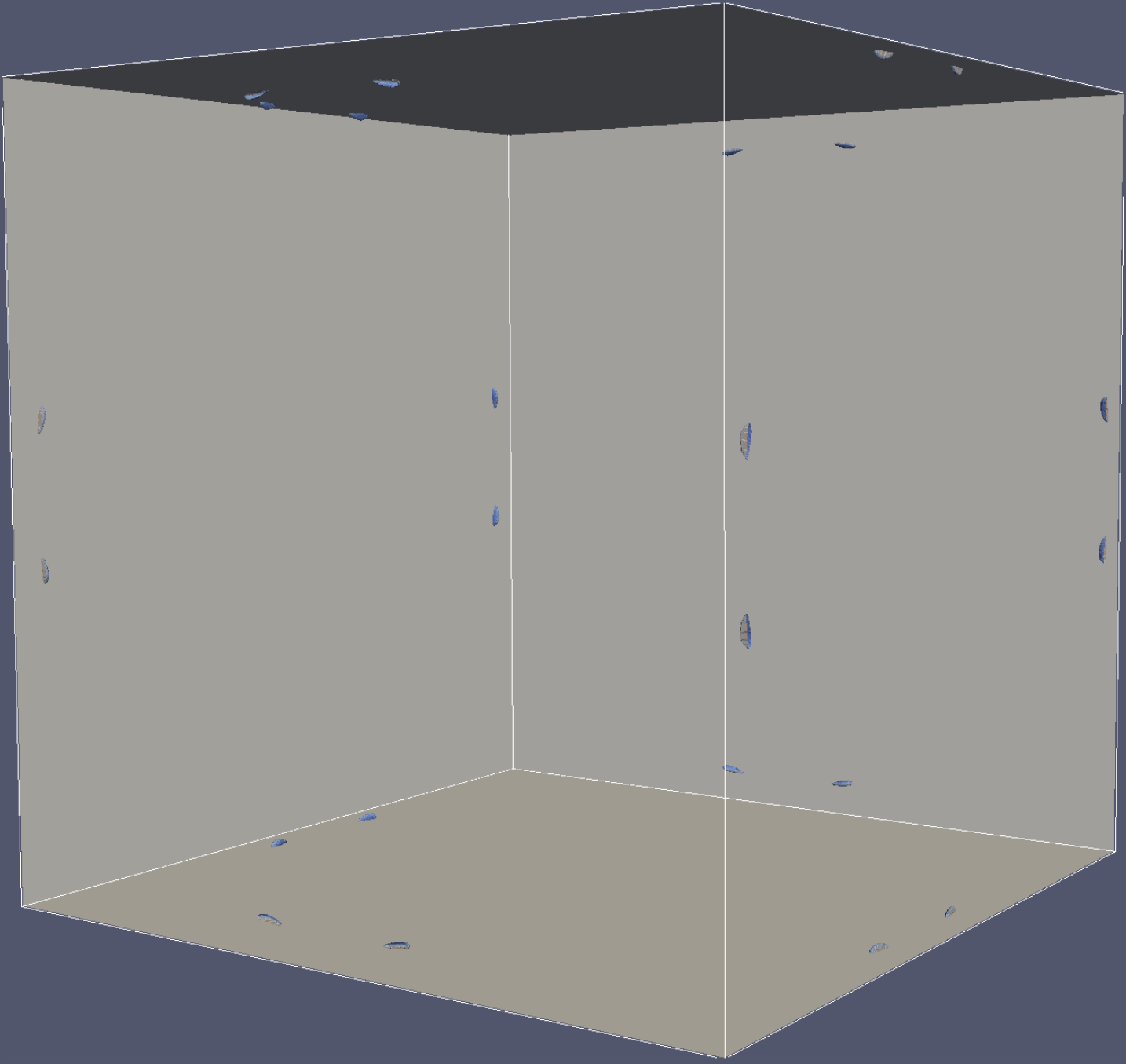}
}
\subfigure[\; $n=1$ , $t=88 000$]{
\includegraphics[width=3.0in]{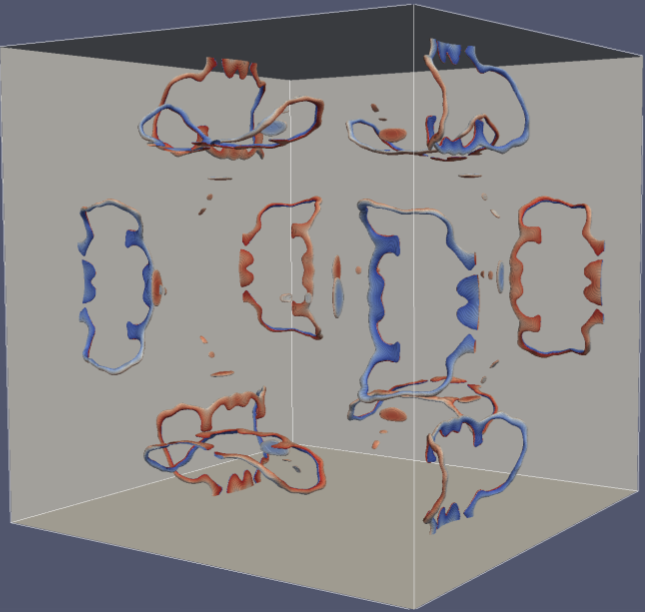}
}
\caption{\label{isosurface_intermittency} \footnotesize 
4 snapshots of the  vortex core singularity isosurfaces at times (a) t = 78 000, (b) t = 81 000, (c)  t = 82 000, and (d) t  = 88 000. Initial conditions of $12$ line vortices.  The phase information (blue is $\phi = 0$, red is $\phi = 2 \pi$ on the vortex core singularities clearly shows the $2 \pi$ phase change in circumnavigating the vortex core.  Grid $1200^3$
 }
\end{center}
\end{figure*}

If one examines the vortex core isosurfaces, Fig. \ref{isosurface_intermittency}, one sees that the topological structure of the vortices are minimal and thus do not permit vortex-wave interactions along these 'point-like' cores. 

\begin{figure*}[!h!t!b!p]
\begin{center}
\subfigure[\; $n=1$ , spectra for $78 000 < t < 86 000$]{
\includegraphics[width=3.15in]{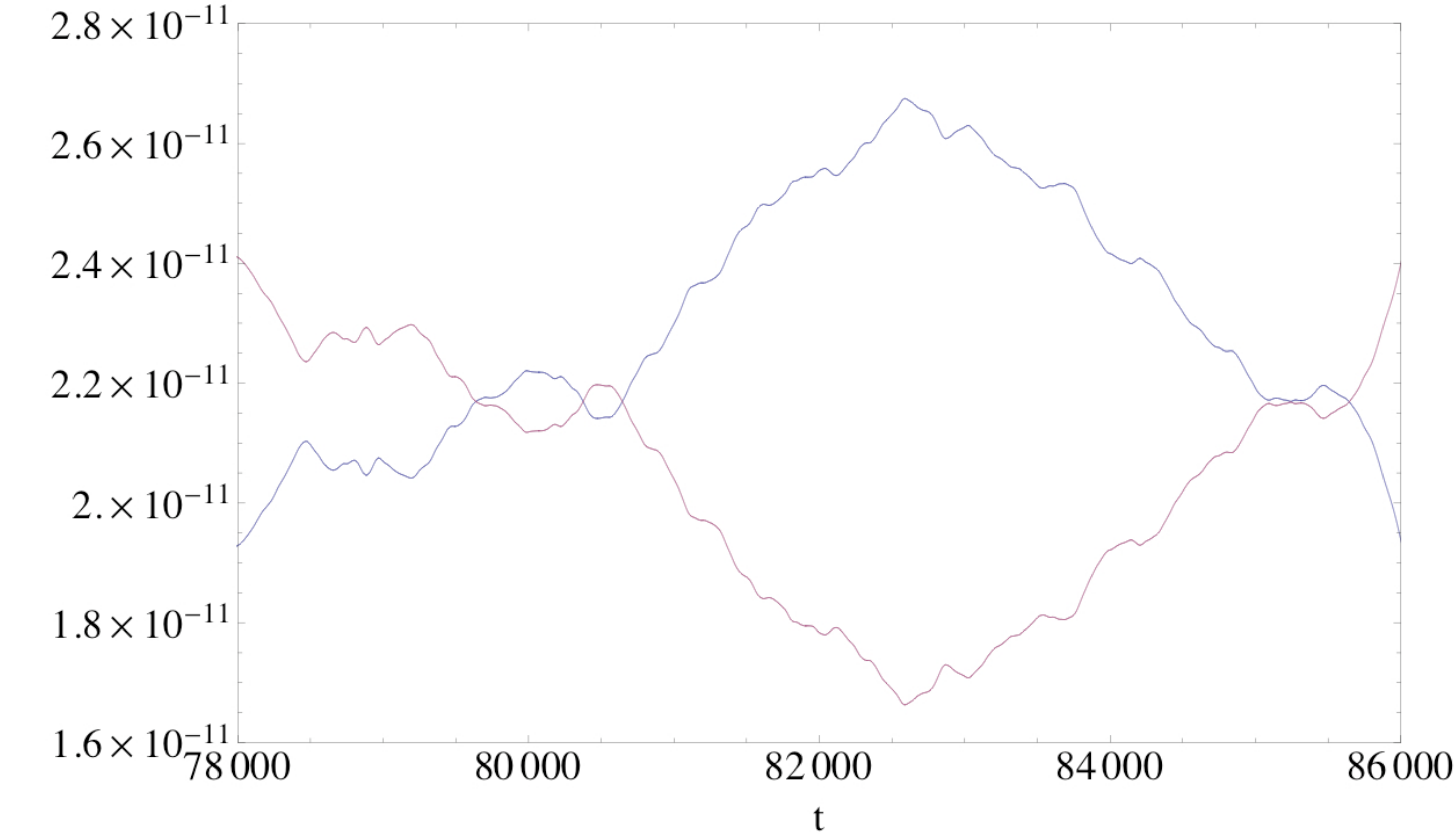}
}
\subfigure[\; $n=1$ , spectra for  $106 000 < t < 120 000$]{
\includegraphics[width=3.15in]{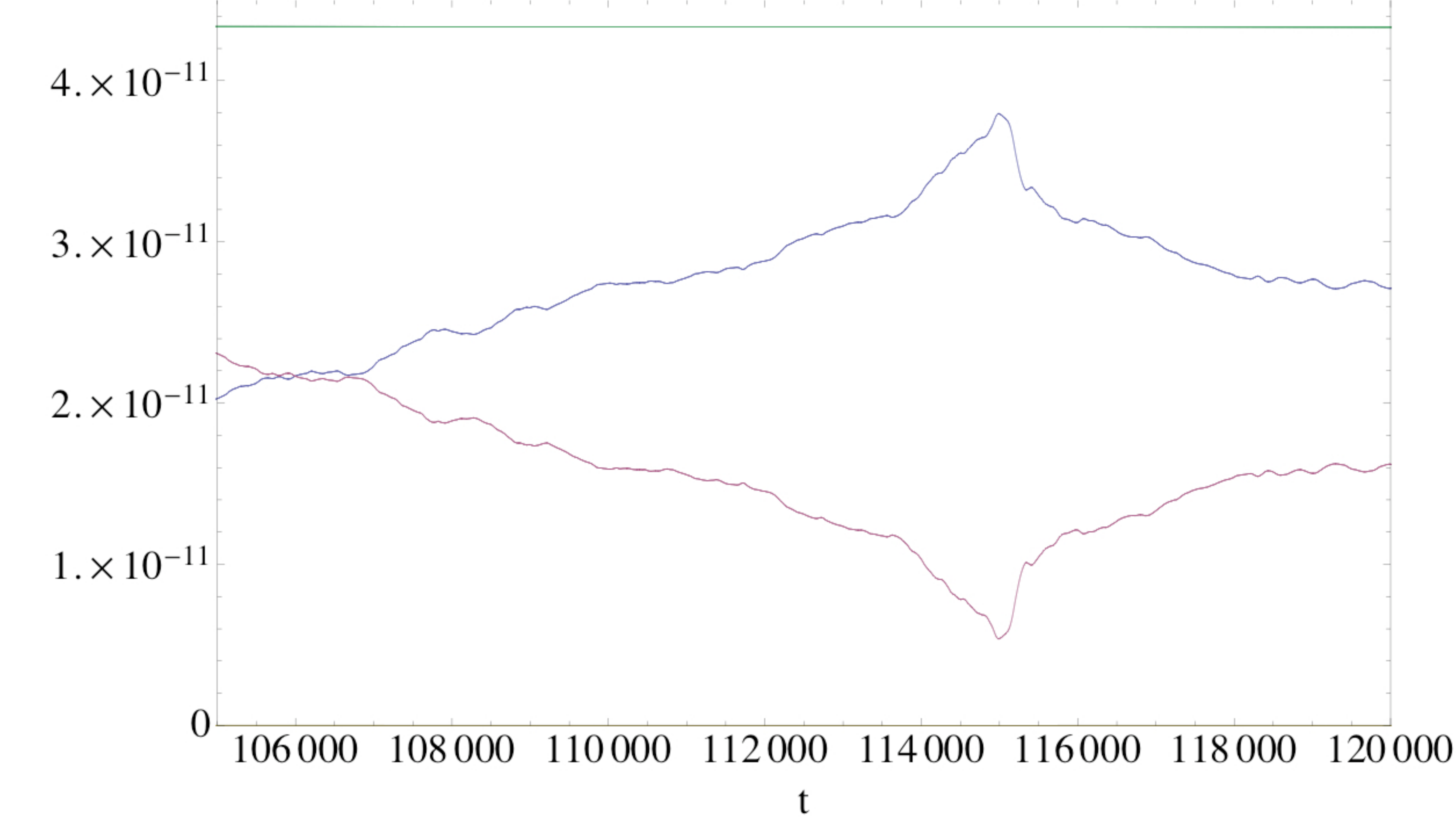}
}
\caption{\label{blow_up_energy} \footnotesize 
Segments in the time evolution of the total kinetic energy $E_{kin}(t)$ (blue) and quantum energy $E_{qu}(t)$ (red) around (a) the loss of vortex cascade around $81 400 < t < 84 300$, and (b) the semi-Poincare recurrence $T_{P}/2 \simeq 115000$. Grid $1200^3$
 }
 \end{center}
\end{figure*}

It is interesting to note that the loss of the quantum vortex $k^{-3}$ spectrum, as seen in the spectral change in the incompressible energy, can also be discerned from the time evolution of the kinetic and quantum energies, Fig. \ref{energy_evolution}.   Indeed, in Fig. \ref{blow_up_energy}, we plot the detailed time evolution of Fig. \ref{energy_evolution} around the loss of vortex cascades $(81 400 < t < 84 300)$ and on the onset of the half-Poincare recurrence time $T_{P}/2 = 115000$.   
As is also evident from Fig. \ref{energy_evolution}  and from simulations results, this loss of the $k^{-3}$ spectrum reoccurs in the time interval $196 400 < t < 199 300$ with the full Poincare recurrence of the initial conditions at $T_{P} = 230 000$.  However only the Poincare recurrence time carries over to the case of initial line vortices with winding number $n=2$ - the loss of the vortex cascade is not seen as expected from the split in the degeneracy of the winding number 2 vortices (see Fig. \ref{semi_Poincare_plot}b and Fig. \ref{energy_evolution} b) and the time evolution of $E_{kin}(t)$ and $E_{qu}(t)$ around $(81 400 < t < 84 300)$.  The vortex core isosurfaces   are shown in Fig. \ref{blow_up_energy} at $t=1000$ to illustrate the degeneracy splitting of the winding number $n=2$ vortices and at time $t=82 000$ where some minimal vortices can be seen among those which can support Kelvin wave cascades.
%
%
%
This strong minimalization of the vortex core isosurfaces at the loss of the vortex cascade is also seen at other grid resolution runs, with the time of occurrence scaling with diffusion ordering.  However, it is not that surprising that this loss of the vortex cascades is not seen for higher winding numbers or for more complex initial conditions (e.g., like the 48-vortex case which still retains considerable vortex cores and loops at $t=84000$)

One possible explanation for these spectral results is based on the role of the $k^{-3}$ spectrum.  If one considers an isolated quantum line vortex, its kinetic energy is all incompressible and this incompressible kinetic energy spectrum is, for all wave numbers, $k^{-3}$ \cite{nore1997physfluid}.  When we turn to the dynamics of quantum vortices we have found this $k^{-3}$ spectrum for incompressible and compressible kinetic energy spectra as well as for the quantum energy spectrum for very large $k$.  This $k^{-3}$ spectrum for the incompressible kinetic energy spectrum is intermittently lost when vortex loops are no longer present in the turbulence, as seen in Fig. \ref {loss_incomp_spectrum}.  For more complex initial conditions, like winding number $2$ vortices, the isosurfaces in the time interval $82 200 < t < 84 000$ will retain vortex loops and the incompressible kinetic energy spectrum retains its $k^{-3}$ spectrum.  A very similar result has also been found in 2D GP turbulence simulations, where under initial random phase conditions and constant density (i.e., when there are initially no [point] vortices), many vortices are rapidly born.  One again encounters both very short Poincare recurrence time $T_P$ of that initial random phase as well as point inversion of these phases at $T_P/2$ -- and during these times of no point vortices the $k^{-3}$ incompressible kinetic energy spectrum is lost. 

There is a wave number $k_{\xi}$ at which there is a spectral break in the incompressible kinetic energy spectrum, where for $k < k_{\xi}$ has a spectral exponent 
$\alpha \sim 3.7$ while for $k_{\xi} < k$ the exponent is $\alpha = 3$.  Moreover, at this wave number $k_{\xi}$, we also see the strong break in the compressible and quantum energy spectrum from the semi-classical regime with strong $\alpha > 6$ to the ubiquitous $\alpha = 3$.  It is thus tempting to associate the wave number spectrum $k_{\xi} < k$ with that of an isolated vortex - together with sound/shock waves with sharp density variations so that the compressible and quantum energy spectra have a $k^{-3}$.  The sharper spectra for all the energies in the region $k < k_{\xi}$ could possibly be attributed to either kelvin waves on the quantized vortices or a Saffman-like $k^{-4}$ spectrum due to vorticity discontinuities.  On grids $1200^3$ the coherence wave number  $k_{\xi} \sim 70$, while on grids $3072^3$ the $k_{\xi} \sim 300$, from the simulations presented in the Appendix.

It is also very tempting to associate the small $k$-region in the total energy spectrum with the classical Kolmogorov-like spectrum.  It is important to note that this $k^{-5/3}$ spectrum is \emph{not} seen in the incompressible kinetic energy spectrum - but in the total (and basically the compressible) kinetic energy spectrum.  It is interesting to see that in classical compressible turbulence studies  \cite{genin2010}, it is the full subgrid energy spectrum that is required to model the Kolmogorov $k^{-5/3}$ - not the incompressible part of this subgrid energy spectrum.

 \section{Conclusion}
 A novel quantum unitary lattice gas algorithm is devised to solve the time evolution of the ground state wave function of a zero temperature BEC as given by the GP equation.  We introduce 2 qubits per spatial node and concentrate on the 1 body sector.  A particular interleaved sequence of unitary square root of swap and unitary streaming operators act on a 2-spinor state.  Parameters are so chosen that under diffusion ordering the zeroth moment of the 2-spinor state reduces, in the continuum limit, to the scalar wave function given by the GP equation.  We find a particular set of initial conditions for which the Poincare recurrence is surprisingly short for this Hamiltonian system.  Since the Poincare recurrence time scales with diffusion ordering, as verified by our simulations on the 2-spinor state, we have run only to grids of $1200^3$.  It is also seen that the \emph{compressible} kinetic energy spectrum exhibits 3 distinct power law cascades with the small k cascade corresponding to the Kolmogorov cascade of classical fluid turbulence while the large k spectrum corresponds to the vortex cores themselves.  The \emph{incompressible} kinetic energy spectrum exhibits predominantly a dual cascade with spectra $k^{-3.37}$ for small $k$ and $k^{-3.0}$ for large $k$.  It is interesting to note that recently Kerr \cite{kerr2010} has also reported on a $k^{-3}$ spectrum:  in particular, Kerr considered the reconnection of two antiparallel quantum vortices.  Following the first reconnection, vortex waves propagate along the quantum vortex and increase in amplitude so that secondary reconnection occurs with the formation of vortex rings.  At this stage Kerr \cite{kerr2010} finds an energy spectrum of $k^{-3}$ with the interaction energy concentrated around the vortex cores where $\rho \sim 0$.  We introduce this work to stress that a $k^{-3}$ spectrum may not necessarily imply the spectrum from a simple isolated vortex which indeed does also have a $k^{-3}$ spectrum.  In our simulations, the intermittent loss of the vortex cascade is readily seen in the loss of the $k^{-3}$ spectrum of the incompressible kinetic energy spectrum as well as in the topological minimilization of the vortex cores.
 
 This work was partially supported by the Air Force Office of Scientific Research and the Department of Energy.  Computations were predominantly performed on the SGI Altix ICE at the ARL and ERDC DoD High Performance Computing Centers, with some computations also performed at DoE NERSC facility.

\bibliographystyle{unsrt}

\bibliography{Kelvin}

\section{Appendix :  Quantum Turbulence Spectra on $3072^3$ grids}
Here we consider some detailed spectral simulations on $3072^3$ grids, for both winding number $n=1$ and $n=2$ straight line vortices.  These simulations were run to time $t_{max}=48 000$.  For $n=1$ vortices, $E_{kin}(0)/E_{qu}(0) = 14.0$, while for $n=2$ vortices $E_{kin}(0)/E_{qu}(0) =34.7$ initially.  By about $t = 10 000$ this ratio between the total kinetic and quantum energies has asymptoted to $\sim 1$, with larger fluctuations for $n=1$ winding number vortices.  

\begin{figure*}[!h!t!b!p]
\begin{center}
\subfigure[\; $ n=1$]{
\includegraphics[width=3.0in]{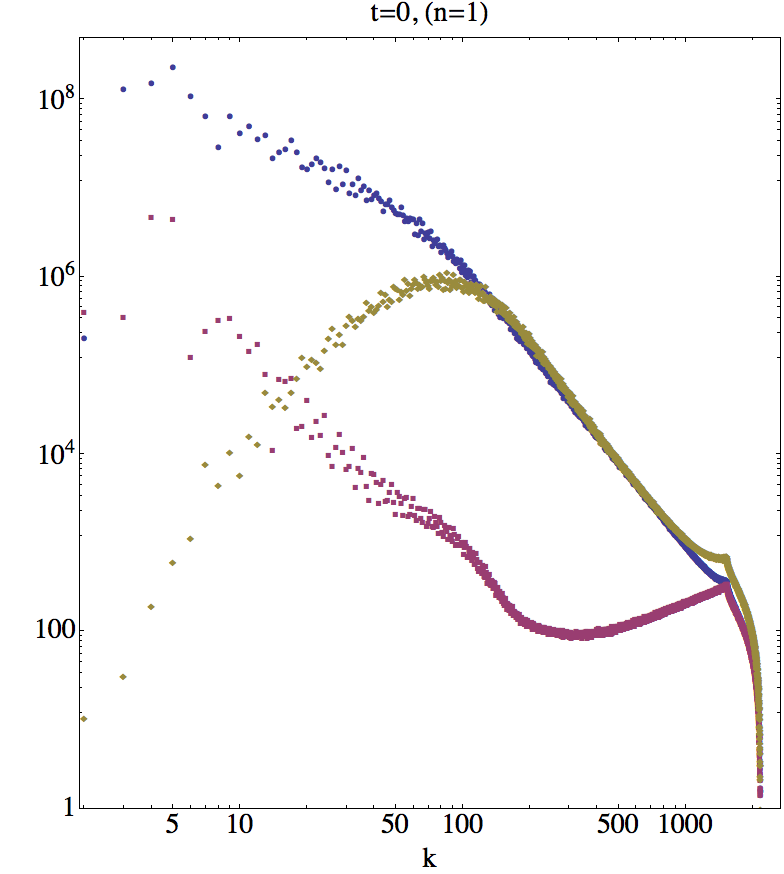}
}
\subfigure[\; $n=2$]{
\includegraphics[width=3.0in]{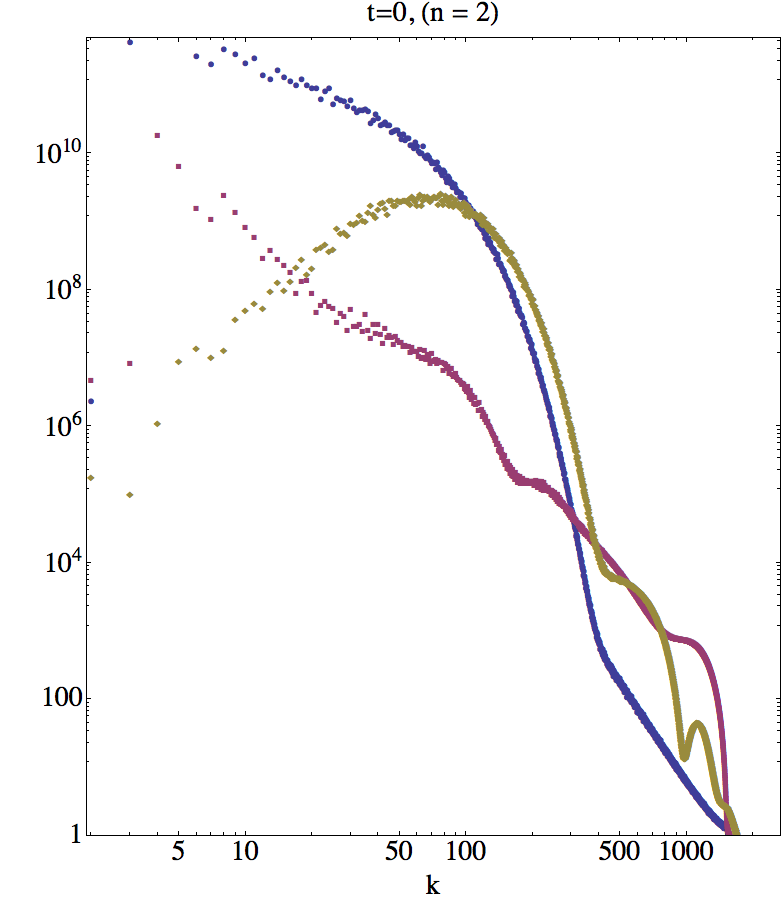}
}
\caption{\label{initial_G3072_spectrum} \footnotesize 
Initial energy spectra for (a) winding number $n=1$ vortices, and (b) winding number $n=2$ vortices.   Blue circles - incompressible kinetic energy, red squares - compressible kinetic energy, gold diamonds - quantum energy. Grid $3072^3$
 }
 \end{center}
\end{figure*}

In breaking down the initial kinetic energy into its compressible and incompressible components.  Initially nearly all the  kinetic energy is \emph{incompressible}
\begin{equation}
n=1 :  E_{kin}^{comp}(0)/E_{kin}^{incomp}(0) = 0.0076,
\end{equation}
\begin{equation}
n=2 :  E_{kin}^{comp}(0)/E_{kin}^{incomp}(0) = 0.0052
\end{equation}
For winding number $n=1$, the compressible energy increases quite rapidly and becomes equal to the incompressible kinetic energy around $t = 11 000$, while for $n=2$ vortices $ E_{kin}^{comp} \sim E_{kin}^{incomp}$ by $t=4000$.  By $t=48 000$, for $n=1$ $E_{kin}^{comp}/E_{kin}^{incomp} \sim 2.3$ but the state is still evolving, while for the more turbulent $n=2$ vortices, this ratio reaches steady state by $t=20 000$ and then fluctuates gently around $E_{kin}^{comp}/E_{kin}^{incomp} \sim 4.0$.  The initial spectra are plotted in Fig. \ref {initial_G3072_spectrum} for quantum, incompressible and compressible energies.

\begin{figure*}[!h!t!b!p]
\begin{center}
\subfigure[\; $ n=1$]{
\includegraphics[width=3.0in]{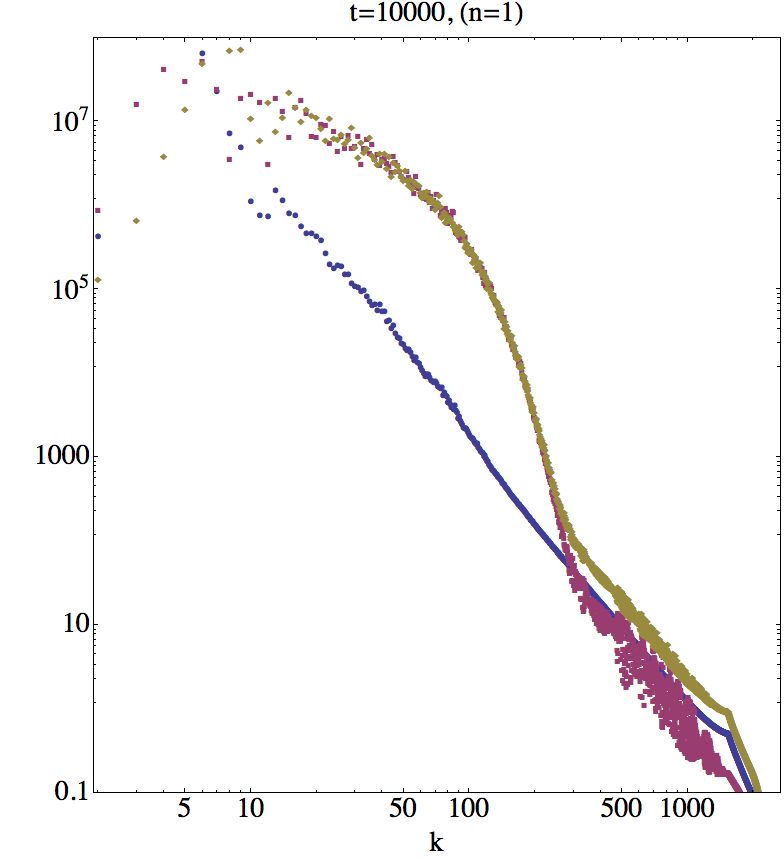}
}
\subfigure[\; $n=2$]{
\includegraphics[width=3.0in]{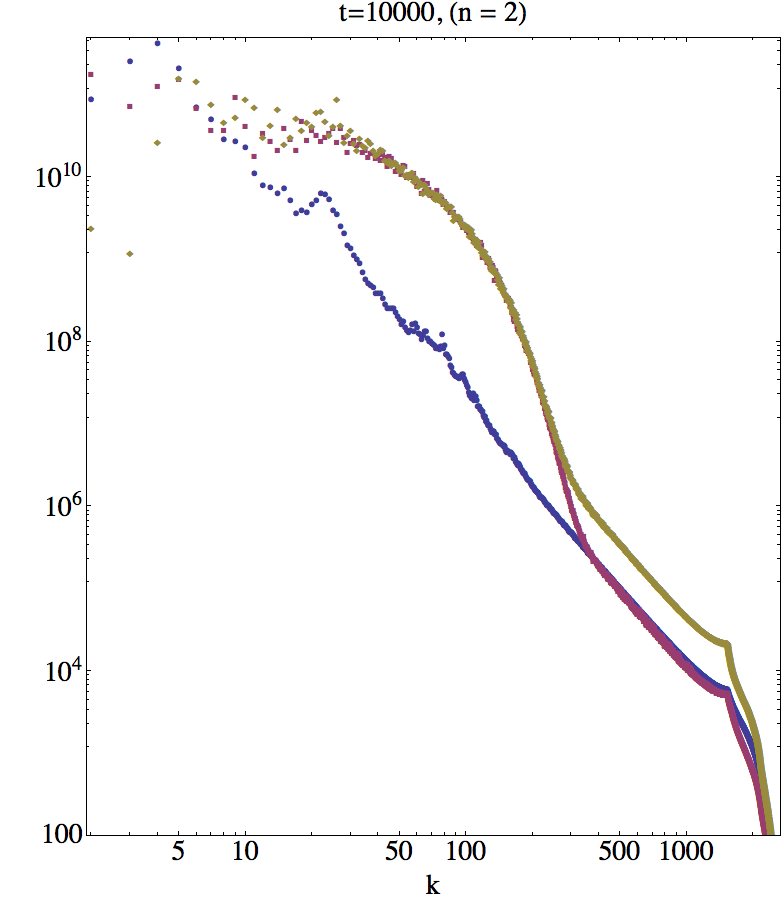}
}
\caption{\label{t10K_G3072_spectrum} \footnotesize 
Energy spectra at $t=10 000$ for (a) winding number $n=1$ vortices, and (b) winding number $n=2$ vortices.   Blue circles - incompressible kinetic energy, red squares - compressible kinetic energy, gold diamonds - quantum energy.  The secondary peak at $k \sim 40$ for $n=2$ vortices in the incompressible kinetic energy spectra appears to be like a backward pulse propagating from much larger $k$ at earlier times.  Grid $3072^3$
 }
 \end{center}
\end{figure*}

\begin{figure*}[!h!t!b!p]
\begin{center}
\subfigure[\; $ n=1$]{
\includegraphics[width=3.0in]{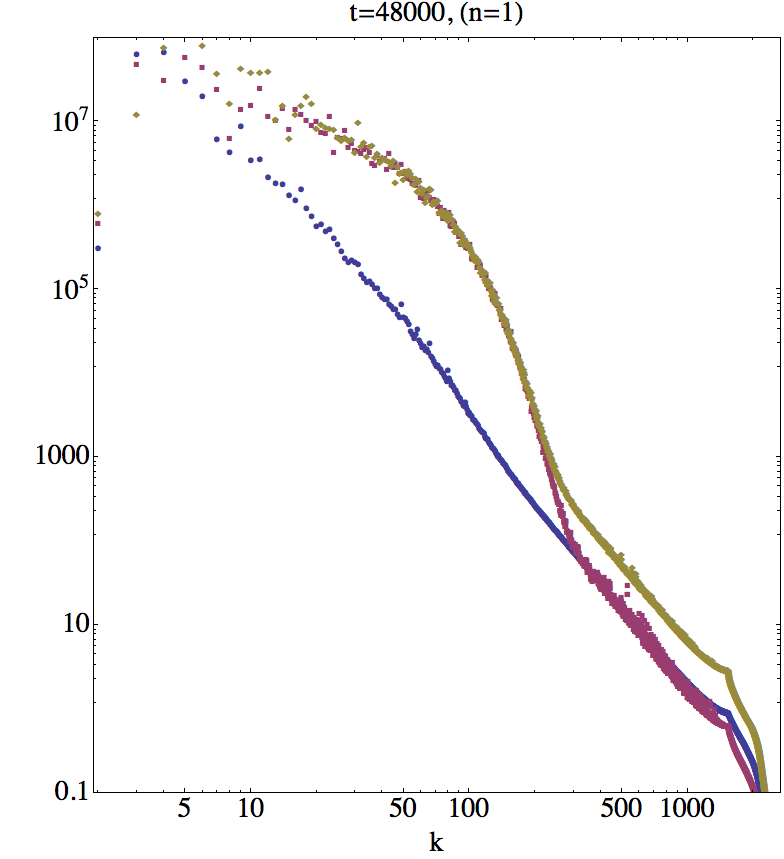}
}
\subfigure[\; $n=2$]{
\includegraphics[width=3.0in]{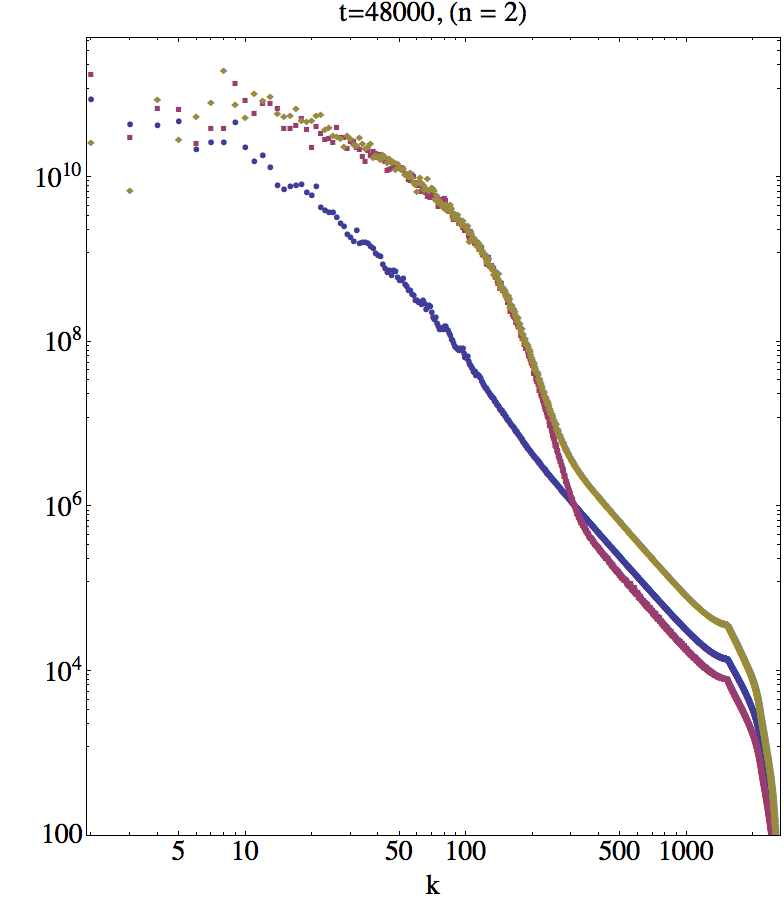}
}
\caption{\label{t48K_G3072_spectrum} \footnotesize 
Energy spectra at $t=48 000$ for (a) winding number $n=1$ vortices, and (b) winding number $n=2$ vortices.   Blue circles - incompressible kinetic energy, red squares - compressible kinetic energy, gold diamonds - quantum energy.  A triple cascade is quite evident in both the quantum and compressible kinetic energy spectra.  These two spectra only deviate around the transition from the medium $k$ to large $k$ cascade, i.e., around $k \sim 300$.  Grid $3072^3$
 }
 \end{center}
\end{figure*}

By $t = 10 000$, the quantum energy spectrum and compressible kinetic energy spectrum are quite similar, Fig. \ref{t10K_G3072_spectrum}, especially for $k > 10$
One is now starting to see the development of a triple cascade spectrum for both the compressible kinetic energy and the quantum energy, while the incompressible energy spectrum has what appears to be a backward pulse propagating (in time) to smaller $k$ from the large $k$ regions.  Moreover, there is the indication that around the wave number $k=300$, at which the compressible and quantum energy spectra enter their third spectral region, there is a distinctt change in the incompressible kinetic energy spectral exponent.  This is further reinforced by examining the spectra at $t=48 000$, Fig. \ref{t48K_G3072_spectrum}.  Note the similarity between the spectra at $t=10 000$ and $t=48 000$.  For winding number $n=1$, the compressible kinetic energy spectrum is quite noisy in the large $k>300$ region, and this noise is considerably suppressed for quantum turbulence driven by winding number $n=2$ vortices.

	Using linear regression in fixed $k$-windows, we compute the time averaged energy spectral exponents and their average deviation for both the $n=1$ and $n=2$ winding number vortices.   The time average is performed from $60$ spectral exponents and their standard deviations determined from times $t = 24 000$ to $t=48 000$ in steps of $\Delta t = 400$. 
\begingroup
\begin{table}[thbp!]
\centering \caption{\footnotesize The time-averaged spectral energy exponents, $k^{-\alpha}$, in the various $k$-bands for winding number $n=1$.  Grid $3072^3$.}   
\label{tab1} \smallskip
\begin{tabular}
{|l|l|l|l|} 
\hline 
{\sc k-band}
&
{\sc Energy}
& 
{\sc $< \alpha>$ }
& 
{\sc $\pm < \sigma>$  }
\\
[0.25ex]
\hline 
$15 < k < 90$ 
& 
Incomp. K.E
& 
3.19
&
0.17
 \\
[0.25ex]
\hline 
$180 < k < 280$ 
&
Incomp. K.E
&
3.134
&
0.016
\\
[0.25ex]
\hline 
$350 < k < 900$
& 
Incomp. K.E
& 
2.9916
&
0.008
 \\
[0.25ex]
\hline
 $15 < k < 90$ 
& 
Quantum
& 
1.95
&
0.23
\\
[0.25ex]
\hline
$180 < k < 280$ 
& 
Quantum
& 
7.810
&
0.087
 \\
[0.25ex]
\hline 
$350 < k < 900$ 
&
Quantum
&
3.054
&
0.060
\\
[0.25ex]
\hline 
$15 < k < 90$ 
& 
Total K.E
& 
1.92
&
0.23
 \\
[0.25ex]
\hline 
$180 < k < 280$ 
&
Total K.E
&
8.606
&
0.084
\\
[0.25ex]
\hline 
$350 < k < 900$ 
& 
Total K.E
& 
3.047
&
0.096
 \\
[0.25ex]
\hline 
\end{tabular}
\end{table}
 \endgroup
%
\begingroup
\begin{table}[thbp!]
\centering \caption{\footnotesize The time-averaged spectral energy exponents, $k^{-\alpha}$, in the various $k$-bands for winding number $n=2$.  Grid $3072^3$.}   
\label{tab1} \smallskip
\begin{tabular}
{|l|l|l|l|} 
\hline 
{\sc k-band}
&
{\sc Energy}
& 
{\sc $< \alpha>$ }
& 
{\sc $\pm < \sigma>$  }
\\
[0.25ex]
\hline 
$15 < k < 90$ 
& 
Incomp. K.E
& 
2.72
&
0.21
 \\
[0.25ex]
\hline 
$180 < k < 280$ 
&
Incomp. K.E
&
3.377
&
0.021
\\
[0.25ex]
\hline 
$350 < k < 900$
& 
Incomp. K.E
& 
3.006
&
0.011
 \\
[0.25ex]
\hline
 $15 < k < 90$ 
& 
Quantum
& 
1.70
&
0.17
\\
[0.25ex]
\hline
$180 < k < 280$ 
& 
Quantum
& 
7.802
&
0.058
 \\
[0.25ex]
\hline 
$350 < k < 900$ 
&
Quantum
&
3.033
&
0.018
\\
[0.25ex]
\hline 
$15 < k < 90$ 
& 
Total K.E
& 
1.66
&
0.17
 \\
[0.25ex]
\hline 
$180 < k < 280$ 
&
Total K.E
&
8.527
&
0.064
\\
[0.25ex]
\hline 
$350 < k < 900$ 
& 
Total K.E
& 
3.042
&
0.020
 \\
[0.25ex]
\hline 
\end{tabular}
\end{table}
 \endgroup

A representative plot is shown in Fig. \ref{Incomp_Comp_G3072_spectra}  for the spectral regions in the total kinetic energy at time $t = 48 000$ for winding number $n=2$ vortices, and for the twin spectral regions for the incompressible kinetic energy around $k \sim 300$:
  \begin{figure*}[!h!t!b!p]
\begin{center}
\subfigure[\; $ n=1$]{
\includegraphics[width=3.0in]{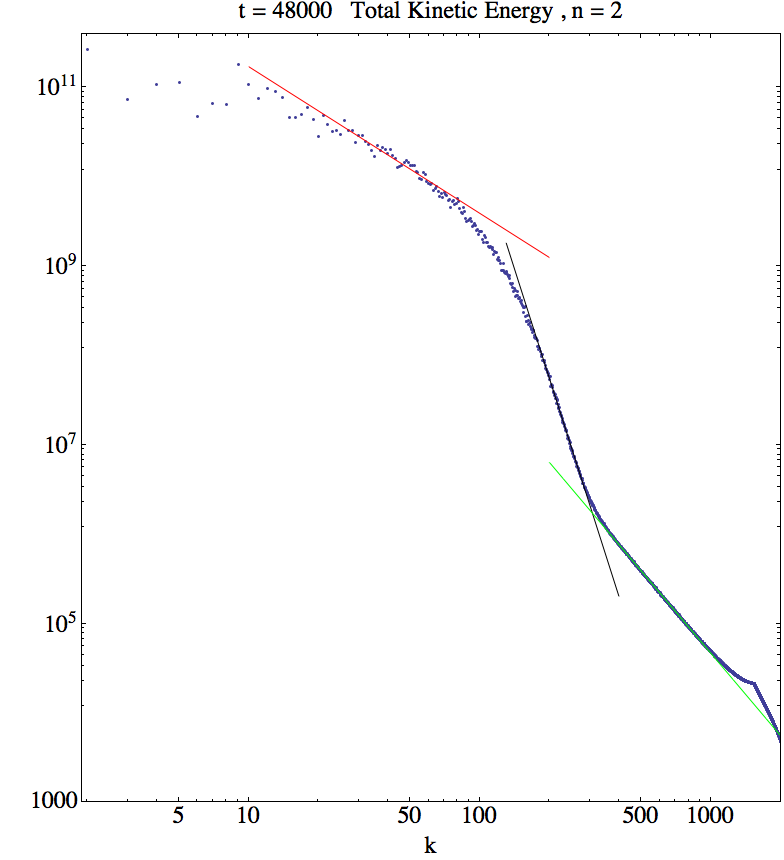}
}
\subfigure[\; $n=2$]{
\includegraphics[width=3.0in]{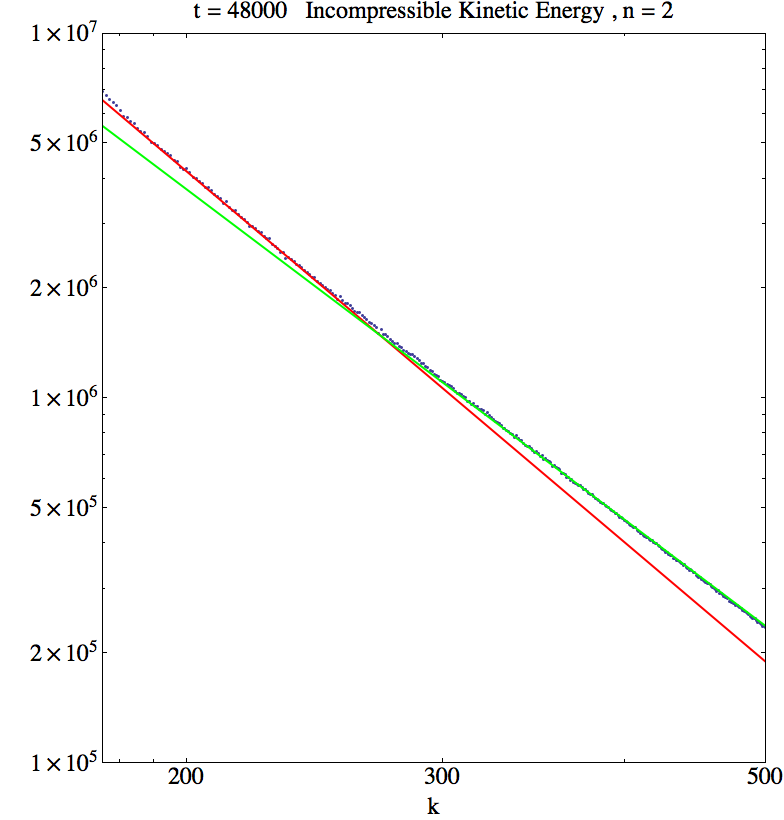}
}
\caption{\label{Incomp_Comp_G3072_spectra} \footnotesize 
Energy spectra at $t=48 000$ for winding number $n=2$ vortices for (a) the total kinetic energy spectrum $E_{kin}(k)$, and (b) the incompressible kinetic energy spectrum $E_{kin}^{incomp}(k)$.   The spectral exponents for the $E_{kin}(k) \sim k^{-\alpha}$ are:  small-$k$ region, $\alpha = 1.64$, for the medium-$k$-range $\alpha = 8.10$ while for the large-$k$ region $\alpha = 3.06$.  For (b) the incompressible kinetic energy spectrum, we show the regression fit for medium-$k$ range with $\alpha = 3.38$ (red line) and large-$k$ region with $\alpha = 3.01$ (green line).  Grid $3072^3$
 }
 \end{center}
\end{figure*}

One can make some conjectures as to physics behinds these exponents.  One possible conjecture is that for large $k > 350$ one is basically looking at the intravortex spectrum of a 'single' vortex as the incompressible kinetic energy spectrum is a very robust $k^{-3}$.  Interestingly, the compressible and quantum energy spectra also exhibit a marked $k^{-3}$ for $k>350$ -- which is not a characteristic of the spectrum of a single linear quantum vortex.  Around $k \sim 350$, both the compressible and quantum energy spectra lift sharply away from the $k^{-3}$ spectrum as we move to smaller $k$.  Around this sharp rise, the incompressible spectrum moves significantly away from $k^{-3}$ to $k^{-3.377}$ for $k < 280$ - at least for the winding number $n=2$ case.  (For $n=1$, the quantum turbulence is more subdued and this could account for the lower exponent of $\alpha = 3.134$ in the incompressible kinetic energy).  Whether this higher exponent is due to quantum Kelvin waves on the vortices needs further investigation.  The small $k$ spectral region is also of some interest.  It is tempting to identify this region with the classical Kolmogorov $k^{-5/3}$ energy spectrum - as the total kinetic energy spectrum of our compressible quantum system exhibits a $k^{-1.66}$ spectrum. for the winding number $n=2$ vortices.  It must be remembered that when dealing with Fourier transforms we have lost all spatial information per se, and so the large $k$- regions are picking up information on physical processes with small spatial scale separation.  While intravortex physics is one such effect, there will be effects from the propagation of wave fronts and shocks since the BEC gas is compressible.

\end{document}